\documentclass[aps,prb,reprint,showpacs,superscriptaddress,amsmath,amssymb]{revtex4-1}

\usepackage{bm}
\usepackage{xcolor}
\usepackage{graphicx}
\usepackage{lineno}
\usepackage{comment}
\usepackage{amsmath}
\usepackage{amssymb}
\usepackage[caption=false]{subfig} 
\usepackage{hyperref}
\usepackage{url}
\usepackage{physics}
\usepackage[normalem]{ulem}

\newcommand{\be}{\begin{equation}}
\newcommand{\ee}{\end{equation}}
\newcommand{\bea}{\begin{eqnarray}}
\newcommand{\eea}{\end{eqnarray}}

\def\rr		{{\bf r}}

\def\QQ		{{\bf Q}}

\def\KK		{{\bf K}}

\def\qq		{{\bf q}}
\def\kk		{{\bf k}}

\def\ga         {\alpha}

\def\go         {\omega}

\def\la         {\langle}
\def\ra         {\rangle}

\renewcommand{\(}{\left(}
\renewcommand{\)}{\right)}

\def\Vscf       {\widehat{V}_{scf} }

\def\bverb      {\renewcommand{\baselinestretch}{1}\begin{verbatim}}
\def\everb  	{\end{verbatim}\renewcommand{\baselinestretch}{1.3}}

\def\ga         {\alpha}

\def\go         {\omega}

\def\la         {\langle}
\def\ra         {\rangle}
\def\kk         {{\bf k}}
\def\qq         {{\bf q}}

\renewcommand{\(}{\left(}
\renewcommand{\)}{\right)}

\newcommand{\ben}{\begin{equation*}}
\newcommand{\een}{\end{equation*}}
\newcommand{\bean}{\begin{eqnarray*}}
\newcommand{\eean}{\end{eqnarray*}}

\renewcommand{\(}{\left(}
\renewcommand{\)}{\right)}

\def\hbn{\textit{h}-BN}

\newcommand{\cinam}{CNRS/Aix-Marseille Universit\'e, Centre Interdisciplinaire de Nanoscience de Marseille UMR 7325 Campus de Luminy, 13288 Marseille cedex 9, France}
\newcommand{\etsf}{European Theoretical Spectroscopy Facilities (ETSF)}
\newcommand{\cnr}{CNR-ISM, Division of Ultrafast Processes in Materials (FLASHit), Area della Ricerca di Roma 1, Via Salaria Km 29.3, I-00016 Monterotondo, Scalo, Italy}

\newcommand{\cnrmodena}{CNR‐NANO, Via Campi 213a, 41125 Modena, Italy}

\graphicspath{{./}}

\begin{document}
\title{First-principles study of luminescence in hexagonal boron nitride single layer: exciton-phonon coupling and the role of substrate}

\author{Pierre Lechifflart}
\affiliation{\cinam}
\author{Fulvio Paleari}
\affiliation{\cnrmodena}
\author{Davide Sangalli}
\affiliation{\cnr}
\author{Claudio Attaccalite}
\affiliation{\cinam}
\affiliation{\etsf}

\begin{abstract}

Hexagonal boron nitride (hBN) is a wide band gap material with both strong excitonic light emission in the ultraviolet and strong exciton-phonon coupling. Luminescence experiments performed on the recently synthesized monolayer form (m-hBN) present emission spectra that differ from one another, with some suggesting a coexistence between phonon-assisted and direct emission channels. Motivated by these results, we investigated the optical response of (m-hBN) using a new \textit{ab initio} approach that takes into account the effects of atomic vibrations on the luminescence spectra. We construct the dynamical exciton-phonon self-energy, then use it to perturbatively correct the optical response functions and test this approach on bulk hBN as a benchmark. Within our approach we are able to estimate the renormalisation of the direct peak induced by phonon-assisted transitions, and this allows us to accurately describe spectra where both processes are present.
We found that the emission signal of m-hBN is strongly dependent on its interaction with the substrate, which changes its nature from direct to indirect material and modifies the screening felt by the electrons. We attribute the m-hBN emission signal to the bright direct excitons and consider the likelihood of phonon replicas appearing.
\end{abstract}

\maketitle

\section{Introduction}

In recent years, single or few-layer materials have attracted a great deal of attention due to their peculiar properties, often different from their bulk counterparts. For example, MoS$_2$ undergoes an indirect-to-direct band gap transition when reducing its thickness to the monolayer limit.\cite{Splendidiani_2010,Mak_2010}
This transition was discovered thanks to the increase in the luminescence signal, since it is well-known that indirect materials tend to be poor light emitters due to higher-order processes mediating the electron-hole recombination.
A similar band gap transition was also predicted for \hbn.\cite{paleari2018excitons}

For many years it was not possible to measure the luminescence signal of a single hBN layer\cite{schue2016dimensionality}, and this was attributed either to the increase of the exciton-exciton annihilation rate in low-dimensional structures\cite{yuan2015exciton,plaud2019exciton} or to other quenching mechanisms.
However, recent experiments reported a luminescence signal from direct excitons in single-layer hexagonal boron nitride (m-hBN) on graphite, showing the existence of a fine two-peak structure.\cite{elias2019direct,wang2022scalable} These experiments were later repeated using exfoliated hBN on a silicon oxide substrate\cite{rousseau2021monolayer}, where only one dominant peak was found.
Indeed, the various m-hBN spectra that appeared in the literature present notable differences which were attributed first to coupling with phonon modes and later to the presence of bubbles in the m-hBN structure.
In addition, the first luminescence measurement on Bernal-stacked bulk hBN was reported recently.\cite{rousseau2022bernal}
These measurements seemed to show the coexistence of emission peaks from both direct and indirect excitons in the same spectrum.
From a theoretical point of view, m-BN has been always considered a direct band gap materials in models\cite{galvani2016excitons}, while using more \emph{ab initio} approaches the nature of its gap depends on the approximation used in the calculations.\cite{prete2020giant,mengle2019impact} Regarding bulk hBN, models and \emph{ab initio} calculations agree on its nature as an indirect gap insulator.\cite{sponza2018direct}
For the intermediate situation, for few-layers hBN, the magnitude and nature of the quasiparticle band gap depends both on the number of layers and on the stacking order.\cite{sponza2018direct,mengle2019impact,latil2022electronic}

In light of these results, we decided to investigate luminescence in m-hBN using a novel approach that includes the coupling between excitons and phonons within an \emph{ab initio} framework and allows for an accurate treatment of both direct and phonon-assisted peaks in the spectra.
The motivation of this study is twofold.
First, m-hBN could present both direct and indirect peaks in its luminescence spectra, which is an ideal test for our theory, while its well-known bulk counterpart provides an excellent benchmark.
Second, the presence of new and partially unclear experiments makes the application of this new methodology interesting and timely.

In condensed matter, the problem of exciton-phonon coupling and phonon-assisted luminescence is an old topic. The first studies date back to the $60$s by Toyozawa \emph{et al.}\cite{toyozawa2003optical,toyozawa1964interband} and the first dynamical solution of the Bethe-Salpeter equation (BSE), the so-called Shindo solution, was proposed precisely to study the exciton-phonon problem.\cite{shindo1970effective}
More recently, this problem has been studied in different materials, from nanotubes\cite{perebeinos} to 2D crystals, and new methodologies were introduced, such as the cumulant Ansatz\cite{cudazzo2020first}, polaron transformation,\cite{feldtmann2009phonon} density matrix,\cite{brem2020phonon} two-particles Green's functions\cite{antonius2017theory} and real-time approach,\cite{paleari2022coupling} with the aim of deriving a modern formulation of exciton-phonon interaction and its related observables, such as phonon-assisted luminescence, within many-body perturbation theory (MBPT).
At variance with older theoretical works, where the values of the exciton-phonon coupling matrix elements were taken as parameters, recent formulations focused on accurate \emph{ab initio} numerical simulations, either tackling the exciton-phonon problem by means of finite-difference displacements in supercells~\cite{cannuccia2019theory,paleari2019exciton}, or, more recently, by combining density functional perturbation theory (DFPT) with Bethe-Salpeter equation (BSE) simulations, in order to avoid the need of large supercells.\cite{chen2020exciton}
In the present work we put forward a formal derivation within MBPT which captures exciton-phonon mediated photoluminescence in a steady-state approximation, and combine it with DFPT to perform accurate \emph{ab initio} numerical simulations in the primitive unit cell.
The great advantage of this formulation is the possibility of integrating over exciton momenta in the full Brillouin zone, thus calculating the renormalisation of the direct peak induced by the indirect transitions. This is essential when studying an emission spectrum that may have competing direct and indirect peaks, such as the case investigated here.

This paper is organized in the following sections: in Sec.~\ref{theory} we describe the theory and the corresponding approximations, followed by the computational approach and relative parameters; in Sec.~\ref{results} we present results for bulk hBN and m-hBN along with a detailed comparison with the different experimental measurements; finally, in Sec.~\ref{conclusions} we draw conclusions and discuss perspectives and future developments.

\section{Theory}\label{theory}

In this section we derive the equations for phonon-assisted absorption and emission based a Dyson-like equation for interacting excitons, where the interaction is mediated by phonons.
We treat "non-interacting" excitons as boson-like composite particles, bound by the Coulombian interaction. This approximation has been shown to correctly reproduce several experimental results\cite{perebeinos,paleari2019exciton} and is expected to be valid in the limit of low excitonic density.
The possibility of treating the electron-electron and electron-phonon interactions on the same footing when deriving the excitonic linear response has been examined in some works,\cite{cudazzo2020first,paleari2022coupling} but we will not consider it in the present formulation. Furthermore, we will be only concerned with phonon-induced corrections up to first order (e.g., we will not describe multi-phonon replicas in luminescence spectra or exciton energy renormalisations\cite{adamska2021bethe,Filip2021phononscreening}).

\subsection{Exciton-phonon coupling}

In order to obtain correlated electron-hole excitations, the so-called excitons, we solve the BSE in the static ladder approximation.\cite{strinati}
The Bethe-Salpeter equation (BSE) reads:
\bea
L(1234) &=& L_0(1234) \nonumber \\
&+& \int d5678 \ L_0 (1256)\  [ v(57)\delta_{56}\delta_{78} \nonumber \\
&-& W(56)\delta_{57}\delta_{68} ] \ L(7834) \label{dyson0}
\eea
where $(1)$ is a shorthand notation for position, time, and spin $(\rr_1, t_1,\sigma_1)$, L is the two-particle correlation function, $L_0$ is its independent-particle version. The terms $v$ and $W$ are, respectively, the bare repulsive and statically-screened attractive Coulomb potentials, derived from the Hartree and screened exchange parts of the electron-electron interaction.\footnote{In the Dyson Eq.~\eqref{dyson0} we replace $v$ with $\bar v$, i.e. the Coulomb potential minus its long-range tail, so as to consider only ``longitudinal'' excitons.\cite{bussi2004effects} We verified that our results for hBN are not affected by this approximation.}
The Dyson equation in Eq.~\eqref{dyson0} can be formally inverted as $L=[\omega - H_{exc}]^-1$, meaning that it can be reformulated\cite{strinati,bussi2004effects} as an eigenvalue problem in the electron-hole transition space, with the index $t$ representing the electronic transition $(n_1 \kk-\QQ) \rightarrow (n_2 \kk)$, where the band indices $n_1$ and $n_2$ belong to occupied (v) or unoccupied (c) states only.
The excitonic Hamiltonian in transition space reads:
\be
\langle t|H_{exc}|t^\prime \rangle = E_t \delta_{t, t'}+ \langle t|v - W |t^\prime \rangle
\label{BSE_H}
\ee
where $E_t$ is the energy associated to the transition $t$.
Notice that we have an Hamiltonian for each transferred momentum $\QQ$. Within the Tamm-Dancoff approximation these Hamiltonians are Hermitian and can be diagonalized in the form
\be
H_{exc} (\QQ) A_\lambda (\QQ) = E_\lambda ( \QQ ) A_\lambda (\QQ),
\label{bse_q}
\ee
where $E_\lambda(\QQ)$ are the eigenvalues, namely the excitonic energies, and $A_\lambda(\QQ)$ the eigenvectors.
With these ingredients we can reconstruct the two-particle correlation function solution of Eq.~\ref{dyson0} as:\footnote{In the two-particle correlation function we consider only the resonant part, because for a system with such a large band-gap the non-resonant part will give zero contributions.}
\be
L (\omega ; \QQ) = \sum_\lambda | \lambda; \QQ \rangle \frac{1}{\omega - E_\lambda(\QQ) + i\eta} \langle \lambda; \QQ |.
\label{eq:static_excL1}
\ee
where $| \lambda; \QQ \rangle$ are the excitonic states solution of Eq.~\eqref{BSE_H}. Multiplying left and right by the electron-hole transition states $t$ and $t'$, and using  $\langle t; \QQ |  \lambda; \QQ \rangle = A_\lambda^{t} (\QQ)$, the two-particle correlation function $L$ can be rewritten as:
\be
L_{t,t'} (\omega ; \QQ) = \sum_\lambda \frac{A_\lambda^{t} (\QQ) \left[ A^{t'}_{\lambda} (\QQ) \right ]^*}{\omega - E_\lambda(\QQ) + i\eta}.
\label{eq:static_excL}
\ee
From $L$, Eq.~\eqref{eq:static_excL1} and \eqref{eq:static_excL}, we can define the corresponding polarization function $\chi(\omega)$ and macroscopic dielectric constant $\epsilon_M(\omega)$ as:
\bea
\chi(\omega)&=& \sum_\lambda \frac{ 1}{\omega - E_\lambda + i\eta} = \sum_\lambda \chi_\lambda (\omega),  \label{eq:chi} \\
\epsilon_M(\omega)&=1 - & \hat d \chi(\omega) \hat d = 1 - 4\pi\sum_\lambda \frac{ | T_\lambda|^2}{\omega - E_\lambda + i\eta},
\eea
where $T_\lambda =  \sum_{cv\kk} d_{cv\kk} A_\lambda^{(cv\kk)}$ are the excitonic dipoles and $d_{cv\kk}$ are the dipole matrix elements between the Kohn-Sham states $d_{cv\kk} = \langle v \kk | \hat r | c\kk \rangle$.

We now introduce the coupling with phonon modes that will mix excitons belonging to different branches and with different momenta.We start from the electron-phonon coupling matrix elements that are obtained from DFPT in the form:\cite{giustino2017electron}
\bean
g_{n n',\mu}(\kk,\qq)=\sum_{s \ga} \frac{e^{i\qq\cdot\tau_s}}{\sqrt {2 M_s \go_{\qq \mu}}}  \xi^{\ga,s}_{\qq \mu}
 \la n\kk |\frac{\partial \Vscf\(\rr\)}{\partial{R_{s\ga}}} | n' \kk-\qq \ra \label{eq:gkkp}
\eean
where $M_s$ is the mass of the $s$-th nucleus, and $\xi^{\ga,s}_{\qq \mu}$ is the polarization vector of the lattice vibration corresponding to wavevector $\qq$ and mode $\mu$, and the bra-ket is the average between two states of the Kohn-Sham potential derivative with respect to the displacement of the $\alpha$ coordinate of the $s$ nucleus.
Using these matrix elements we can construct the standard electron-phonon Hamiltonian.\cite{giustino2017electron} Then Green's function theory can be used to construct the response functions in presence of the electron-electron plus electron-phonon interaction. This procedure gives rise to a dynamical Bethe-Salpeter Equation.\cite{cudazzo2020first,paleari2019first} This equation is not easy to solve due to the presence of two times (or frequencies). However, the problem can be mapped into an exciton-phonon Hamiltonian:
\bea
H_{exc-ph} (\QQ)&=& \sum_\lambda E_\lambda(\QQ) \hat a_{\lambda\QQ}^{\dagger} \hat a_{\lambda\QQ} + \sum_{\mu \qq} \hbar \omega_{\mu\qq} \hat b^\dagger_{\mu\qq} \hat b_{\mu\qq} \label{Hexcph} \\
&+& \sum_{nm,\mu\qq} \mathcal{G}_{mn,\mu}(\QQ,\qq) \hat a_{m\QQ+\qq}^{\dagger} \hat a_{n\QQ} (\hat b_{\mu\qq}^{\dagger} + \hat b_{\mu -\qq} )  \nonumber
\eea
where $\hat a_{\lambda\QQ}^{\dagger}, \hat a_{\lambda\QQ}$ are the  creation/destruction operators for the excitons and $\hat b_{\mu\qq}^{\dagger},\hat b_{\mu\qq}$ for the phonons and $ \omega_{\mu\qq}$ are the phonon energies.
The Hamiltonian, Eq~.\eqref{Hexcph}, has been derived in different ways in the literature.
One can start from Eq.~\eqref{BSE_H}, taking it as the unperturbed excitonic Hamiltonian, and introduce a perturbation with respect to the lattice displacements, see Supp. Mat. in Ref.~[\onlinecite{chen2020exciton}].
Another possibility is to start instead from Eq.~\eqref{dyson0} and, using Green's function theory, introduce a dynamical term in the electron-hole kernel involving the phonon propagator. Then, the Dyson equation can be split in two by separating the static contributions (due to electron-electron interactions, Eq.~\eqref{dyson0}) from the dynamical ones (due to electron-phonon interactions, see Eq.~\eqref{dynaeq} in the following) and finally remapping it by formal inversion into an Hamiltonian form.
This approach is followed in Ref.~[\onlinecite{cudazzo2020first}].
Finally, the exciton-phonon coupling can also be derived directly from the general electronic Hamiltonian while treating electron-electron, electron-phonon and external field terms on the same footing (as in Ref.~[\onlinecite{paleari2022coupling}]), a procedure that lifts some approximations with respect to the previous cases, but does not introduce relevant changes for the systems investigated here.

In all cases, the exciton-phonon matrix elements that enter the Hamiltonian are the sum of two contributions relative to the coupling of the phonon with either the hole ($h$) or the electron ($e$) in the pair-particle substructure of the exciton:
\bea
\mathcal{G}_{\beta \lambda,\mu}(\QQ,\qq)=\mathcal{G}^h_{\beta \lambda,\mu}(\QQ,\qq)-\mathcal{G}^e_{\beta \lambda,\mu}(\QQ,-\qq)  \label{eq:Gexcphon}
\eea
in terms of electron-phonon coupling and excitonic states they are defined as:\cite{chen2020exciton,paleari2019first,cudazzo2020first,toyozawa2003optical,antonius2017theory}
 \bean
 \mathcal{G}^h_{\beta \lambda,\mu}(\QQ,\qq) &=& \sum_{\substack{\substack{v,v' \\ c,c',\kk}}}A_{\lambda,\QQ}^{v,c,\kk} \left [g_{vv',\mu}(\kk-\QQ,\qq) \delta_{c,c'} \right] A_{\beta,\QQ+\qq}^{v',c',\kk^* }\\
 \mathcal{G}^e_{\beta \lambda,\mu}(\QQ,\qq) &=& \sum_{\substack{v,v' \\ c,c',\kk}}A_{\lambda,\QQ}^{v,c,\kk}  \left [ g^*_{c'c,\mu}(\kk+\qq,\qq) \delta_{v,v'} \right ] A_{\beta,\QQ+\qq}^{v',c',\kk+\qq^* }.
 \eean
These two terms correspond to a rotation of the electron-phonon coupling in the excitonic basis.
This formulation in the excitonic basis has great advantages. First, it allows for treating the coupling with phonons as a perturbation starting from the solution of an unperturbed Bethe-Salpeter Equation. Second, it allows all scattering channels to be expressed as a transition between excitonic states.\cite{antonius2017theory}

Now that we have a new Hamiltonian for excitons and phonons we can proceed to its solution. Unlike the ``unperturbed'' excitonic Hamiltonian, Eq.~\eqref{BSE_H}, in this case we cannot proceed with a direct diagonalization, because the electron-phonon coupling mixes the transitions at different $\QQ$ and therefore the Hilbert space becomes too large. We will therefore use the method of Green's functions (i.e., the second method described above to derive the exciton-phonon interaction) to find an approximate solution to Eq.~\eqref{Hexcph}. We then consider the Dyson equation with the dynamical exciton-phonon kernel:
\footnote{Note that this equation can also be rederived starting from Eq.~\eqref{Hexcph} in the same way that perturbation theory on the Green's function is applied to the electron case:\cite{mahan2013many,giustino2017electron} we can use the excitonic Hamiltonian to generate the equations of motion (EOM) for the excitonic creation-destruction operators. Then we use these EOMs to find a closed equation for the excitonic Green's $\mathcal{L}$ function in terms of a self-energy that will correspond to the dynamical kernel $\Xi$.}
\be
  \mathcal{L}(1234) = L(1234) + L(1625)\ \Xi^{eph}(5867)\  \mathcal{L}(7483),
  \label{dynaeq}
\ee
where repeated indices are integrated upon. Here $L$ is the solution of the static BSE, the same as in Eq.~\eqref{eq:static_excL}, embodying the static Coulombian electron-hole pair interaction.
The dynamical kernel $\Xi_{ph}$ is responsible for the effective phonon-mediated exciton-exciton interactions (involving electron-phonon and hole-phonon components).
As the kernel is dynamical, the latter equation cannot easily be written in terms of a two-time or one-frequency propagator. It is then not possible to invert it without considering a partial re-summation.\cite{marini2003dynamical,Cudazzo2020satellites}
Instead, we consider the dynamical electron-phonon interaction only up to first order, and define $L^{(1)}$ as the solution of Eq.~\eqref{dynaeq} obtained by replacing $\mathcal{L}$ with $L$ on the r.h.s..
Using a first-order solution, we consider the scattering of an exciton with only one phonon at a time (multiple scattering extensions are discussed in the literature\cite{PhysRevLett.101.057401}).
Note that here the knowledge of the static propagator $L$ from Eq.~\eqref{eq:static_excL}, obtained from the solution of the standard BSE, is needed to solve Eq.~\eqref{dynaeq} even at the first order in $\Xi^{eph}(5867)$. Using the relation $\chi(12) = -i L(1212)$, one can obtain the response functions in the excitonic basis including the first-order correction due to the exciton-phonon coupling as:
\be
  \chi^{(1)}_{\lambda\lambda'}(\omega) = \chi_\lambda(\omega) + \chi_\lambda(\omega) \Xi_{\lambda\lambda'}^{eph}(\omega) \chi_{\lambda'}(\omega) \label{chi_one}
\ee
where $ \chi_\lambda(\omega)$ is a short notation for $ \chi_{\lambda\lambda'} (\omega) \delta_{\lambda,\lambda'}$, see Eq.~\eqref{eq:chi}.

The $\chi$ on the right-hand side is the one obtained with static BSE Eq.~\eqref{eq:chi} and  $\Xi^{eph}$ is the dynamical exciton-phonon self-energy describing the coupling between exciton and phonons.

The self-energy $\Xi^{eph}$ can be explicitly computed via MBPT,\cite{mahan2013many} similarly to how it is done in the electron-phonon case.\cite{giustino2017electron} Owing to the fact that it is at first order in the exciton-phonon scattering, it results in a structure similar the the Fan-Migdal self-energy for electron-phonon problem, but with the electronic propagator being replaced by the excitonic one: $\Xi^{eph} =  \mathcal{G}^2 D L$, where $D$ is the phonon propagator, $L$ the excitonic Green's function solution of Eq.~\eqref{eq:static_excL}, and $\mathcal{G}$ the exciton-phonon matrix elements. Following analogous steps to the electronic case\cite{giustino2017electron} we get the expression for self-energy:
\bea
\Xi^{eph}_{\lambda \lambda'}(\QQ; \omega) &=& \frac{1}{\Omega_{BZ}}\sum_{\qq,\mu \beta}  \mathcal{G}_{\beta \lambda,\mu}(\QQ,\qq) \mathcal{G}^{*}_{\beta \lambda',\mu}(\QQ,\qq) \nonumber \\
&\times& \left [ \frac{ 1 - N_\beta(\QQ+\qq) + n_{\qq,\mu} }{\omega - E_{\QQ+\qq,\beta} + \omega_{\qq \mu }  + i\eta } \right . \nonumber \\
&+& \left . \frac{  N_\beta(\QQ+\qq)+ n_{\qq,\mu} }{\omega - E_{\QQ+\qq,\beta} - \omega_{\qq \mu }  + i\eta }
\right], \label{self-energy}
\eea
where $\beta$ and $\mu$ are exciton and phonon band indices respectively, $\Omega_{BZ}$ is the volume of the reciprocal Brillouin zone, $n_{\qq,\mu}$ and $N_\beta(\QQ)$ the temperature-dependent phonon and exciton occupation factors.

\subsection{Phonon-mediated absorption and emission}

The self-energy Eq.~\eqref{self-energy} depends on the excitonic transferred momentum $\QQ$ in a similar way to how the electronic self-energy depends on the electron crystal momentum $\kk$.\cite{giustino2017electron}
Since we are interested in optical properties that depend only on the $\QQ=0$ excitons, we will specialize the self-energy to this case.
Then, we will make two more approximations to Eq.~\eqref{self-energy} in order to make the problem easier to solve.
First, we will disregard the excitonic occupations, since in general they are very small compared to the phononic ones, especially in the kind of near-equilibrium experiments we are interested in.\footnote{For an estimation of the exciton density in BN luminescence experiments, one can have a look at the Supplemental Material of Ref.~\onlinecite{schue2018direct}.}
Second, we will use a diagonal approximation for the self-energy. For a discussion on the form and possible role of off-diagonal self-energy matrix elements, the interested reader can have a look at Refs.~\onlinecite{toyozawa1964interband,paleari2019first}.
We end up with the following expression:
\bea
\Xi^{eph}_{\lambda }(\QQ=0; \omega) &=& \frac{1}{\Omega_{BZ}}\sum_{\qq,\mu \beta}  \mathcal{G}_{\beta \lambda,\mu}(\qq) \mathcal{G}^{*}_{\beta \lambda,\mu}(\qq) \nonumber \\
&\times& \left [ \frac{ 1 + n_{\qq,\mu} }{\omega - E_{\qq,\beta} + \omega_{\qq \mu }  + i\eta } \right . \nonumber \\
&+& \left . \frac{  n_{\qq,\mu} }{\omega - E_{\qq,\beta} - \omega_{\qq \mu }  + i\eta }
\right],\label{self-energy-simple}
\eea
where $\mathcal{G}_{\beta \lambda,\mu}(\qq)$ is a short notation for $\mathcal{G}_{\beta \lambda,\mu}(\QQ=0,\qq)$.
Replacing Eq.~\eqref{self-energy-simple} in Eq.~\eqref{chi_one}, we can write down the phonon-assisted response function at first order in the exciton-phonon coupling as:

\bea
\chi^{(1)}_{\lambda}(\omega)&=& | T_\lambda  |^2 \left \{ \frac{1 - R_\lambda}{\omega - E_\lambda + i\eta} \nonumber \right . \nonumber \\
&+& \left . \sum_{\mu\beta\qq} |\mathcal{D}^{\pm}_{\beta\lambda,\qq \mu}|^2 \frac{1/2 \pm 1/2 + n_{\qq,\mu} }{ \omega - E_{\qq,\beta} \pm \omega_{\qq \mu}  + i\eta } \right \},
\label{eq:chi_lambda}
\eea

where $R_\lambda$ is the renormalisation factor, $ | T_\lambda | ^2$ are the excitonic dipoles and $|\mathcal{D}^{\pm}_{\beta\lambda,\qq \mu}|^2$ are the phonon-assisted coupling strengths.
In this formula, the first term describes the direct transitions, the weight of which is reduced by $R_\lambda$, while the subsequent terms are the satellites appearing at the energy of finite-momentum excitons $E_{\qq,\beta}$ plus or minus one phonon energy. The $\pm$ sign refers to phonon emission or absorption processes.
We can write down explicitly the renormalisation factor and the phonon-assisted dipole moments as

\bea
R_\lambda&=&-\left . \frac{\partial \Pi^{eph}_{\lambda\lambda}(\omega)}{\partial \omega} \right |_{\omega=E_\lambda} \\
| \mathcal{D}^{\pm}_{\beta\lambda,\qq \mu} |^2  &=& \frac{|\mathcal{G}^{exc}_{\beta\lambda,\mu}|^2}{(E_{0,\lambda}- E_{\qq,\beta} \pm \omega_\mu)^2}. \label{ph-ass-dip}
\eea

Notice that the phonon-assisted dipole moments, Eq.~\eqref{ph-ass-dip}, could diverge if a phonon-mode is resonant with the excitonic energy difference. 
However, as pointed out for example by Toyozawa, see Ch.~$10$ in the book Ref.~[\onlinecite{toyozawa2003optical}], this divergence is an artifact of the finite-order perturbation theory. 
The inclusion of higher orders introduces a shift in the excitonic levels and a broadening that eliminates the divergence.

Using Eq.~\eqref{self-energy-simple}, we can also rewrite the renormalisation factor as:
\bean
 R_\lambda &=& \sum_{\qq,\mu \beta}  | \mathcal{G}_{\beta \lambda,\mu}(\qq) |^2 \\
      &\cdot& \left [ \frac{ n_{q,\mu} + 1 }{(E_\lambda - E_{\qq,\beta} + \omega_{\qq \mu } )^2 } + \frac{ n_{\qq,\mu} }{(E_\lambda - E_{\qq,\beta} - \omega_{\qq \mu } )^2} \right ].
\eean
This factor measures how much of the spectral weight is transferred to the satellites.
Using Eq.~\eqref{eq:chi_lambda} we can describe phonon-assisted light absorption including the renormalisation effects of the direct transitions.
In order to apply Eq.~\eqref{eq:chi_lambda} to the light emission process, we employ a steady-state approximation and we obtain the luminescence via the van Roosbroeck-Shockley (RS) relation\cite{van1954photon,bbwill1972,paleari2019exciton}:
\begin{widetext}
\bea
	{I}^{PL}(\omega)&=& \Im \sum_\lambda \frac{| T_\lambda  |^2}{\pi^2 \hbar c^3} \Biggl\{ \omega^3 n_r(\omega) \frac{1 - R_\lambda}{\omega - E_\lambda + i\eta} e^{-\frac{E_{\lambda}-E_{min}}{kT_{exc}}} \nonumber \\
	&+& \left . \sum_{\mu\beta\qq} \omega(\omega \mp 2 \omega_{\qq \mu})^2 n_r(\omega) \left|\mathcal{D}^{\pm}_{\beta\lambda,\qq \mu} \right|^2 \frac{1/2 \pm 1/2 + n_{\qq,\mu} }{ \omega - (E_{\qq,\beta} \mp \omega_{\qq \mu})
	+ i\eta } e^{-\frac{E_{\qq,\beta}-E_{min}}{kT_{exc}}} \right \}, \label{lum_eq}
\eea
\end{widetext}
where $E_{min}$ is the minimum of the exciton dispersion, $n_r(\omega)$ is the refractive index. The parameter $T_{exc}$ is an effective excitonic temperature, used to model excitonic occupations with a Boltzmann distribution. The value of $T_{exc}$ can be estimated from experimental measurements, see Supplemental Material (SM) \footnote{See Supplemental Material at [LINK] for phonon and exciton dispersions of hBN and m-hBN, discussions on nearly-free states, effective excitonic temperature, phonon mode-resolved luminescence spectra and exciton occupation factors. The SM includes Refs. \onlinecite{kammerlander2012speeding,faber2014excited,sohier2017breakdown,wirtz2003ab,sponza2018exciton,schafer2013semiconductor,libbi2022phonon}.}.
The possibility of using an excitonic occupation determined from the underlying single-particle occupations -- thus deviating from a Boltzmann/Bose-Einstein factor and taking explicitly into account the composite nature of the excitons -- has been discussed in some papers.\cite{Pedro2016PL,cannuccia2019theory}
We have numerically tested this latter possibility but found that it leads to unphysical excitonic occupations (see SM).

Equations \eqref{eq:chi_lambda} and \eqref{lum_eq} are the main formulas of the present manuscript that will be applied in the next sections to different BN-based structures, in such a way to validate the present approach and compare with the experimental measurements.

\subsection{Computational details}\label{calc}

\begin{figure}
    \centering
    \includegraphics[width=\columnwidth]{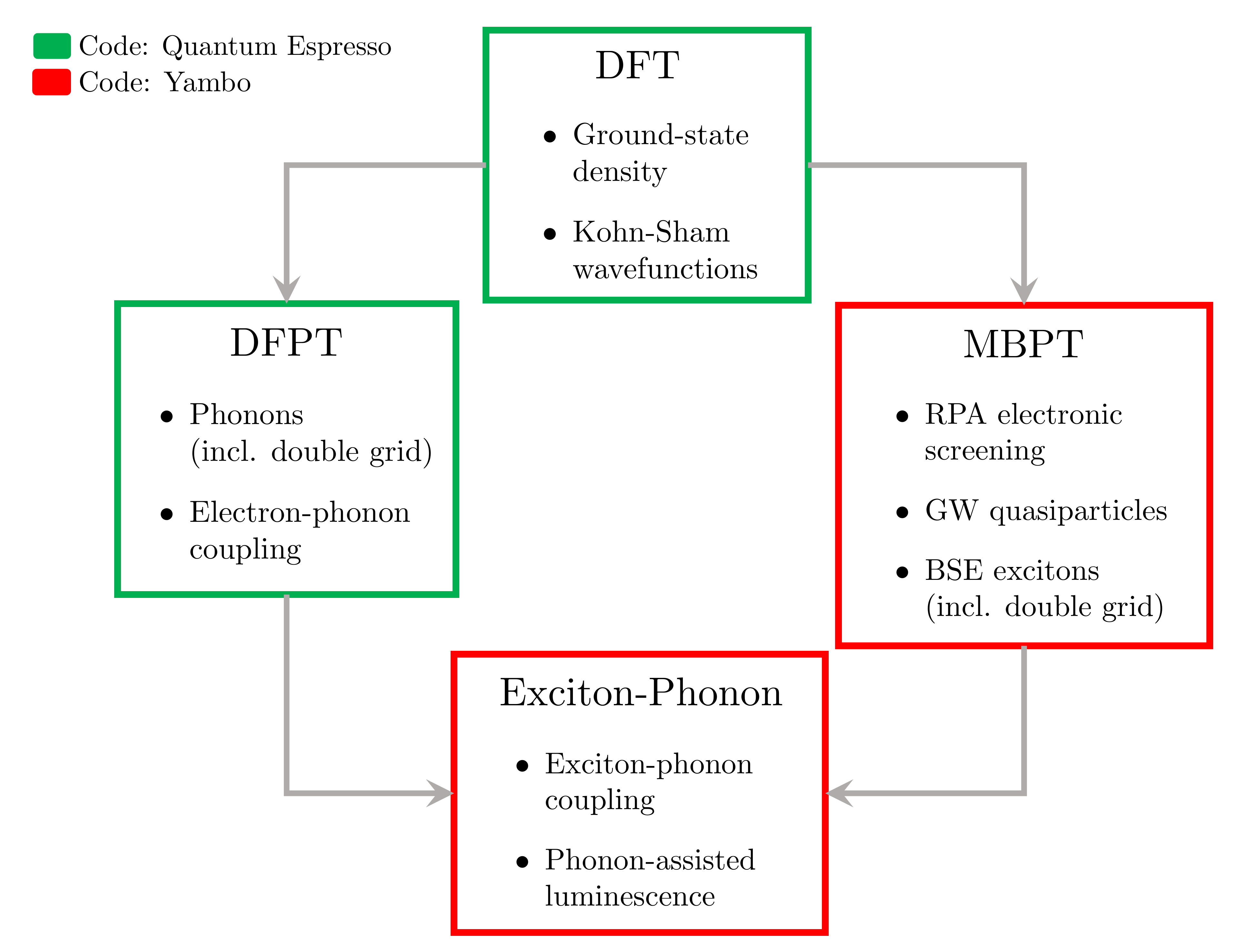}
	\caption{Schematic representation of the \textit{ab initio} workflow to compute exciton-phonon quantities. In this scheme the DFT block includes both the self-consistent calculation for the density and the non-self-consistent part for the wavefunctions. In the DFPT block, phonons and electron-phonon coupling matrix elements are calculated.\cite{pwscf} The MBPT block includes the calculations of dielectric function and quasiparticle band structures and the solution of the Bethe-Salpeter equation.\cite{yambo} The DFPT-MBPT results are finally put together to obtain the exciton-phonon coupling matrix element.\label{fig:scheme} }
\end{figure}

In this section we present our computational workflow for the calculation of the exciton-phonon coupling and all the computational details needed to reproduce the results. In Fig.~\ref{fig:scheme} we report a sketch of the workflow.
We start from the atomic structure, which was optimized in DFT for the monolayer and taken from Ref.~[\onlinecite{sponza2018direct}] for bulk hBN. The relevant lattice parameters are shown in Table~\ref{tab:parms}. We performed all DFT calculations using the Quantum Espresso code\cite{pwscf}, with norm-conserving pseudopotentials\cite{van2018pseudodojo} in the Local Density Approximation (LDA).\cite{PhysRevB.43.1993} In the monolayer case, a supercell was used with a length of 21~\r{A} along the $z$-direction -- in such a way to avoid spurious interactions between periodic images -- along with the 2D cutoff for the Coulomb interaction implemented in Quantum Espresso\cite{sohier2017density}. Energy cutoff and other parameters that enter the DFT calculations are reported in Tab.~\ref{tab:parms}. Phonons and electron-phonon couplings were calculated using DFPT starting from the DFT results. The transferred momenta grid for the phonons is reported in Tab.~\ref{tab:parms}. The correct long-range behavior of the electron-phonon coupling in 2D was obtained by applying a cutoff of the Coulomb interaction in the $z$-direction.\cite{sohier2016two} The electron-phonon matrix elements were calculated on the same $\qq$-grid as the phonons and the excitons, for all the electronic bands entering in the Bethe-Salpeter equation kernel.

\begin{figure*}[ht]
    \centering
    \includegraphics[width=0.98\textwidth]{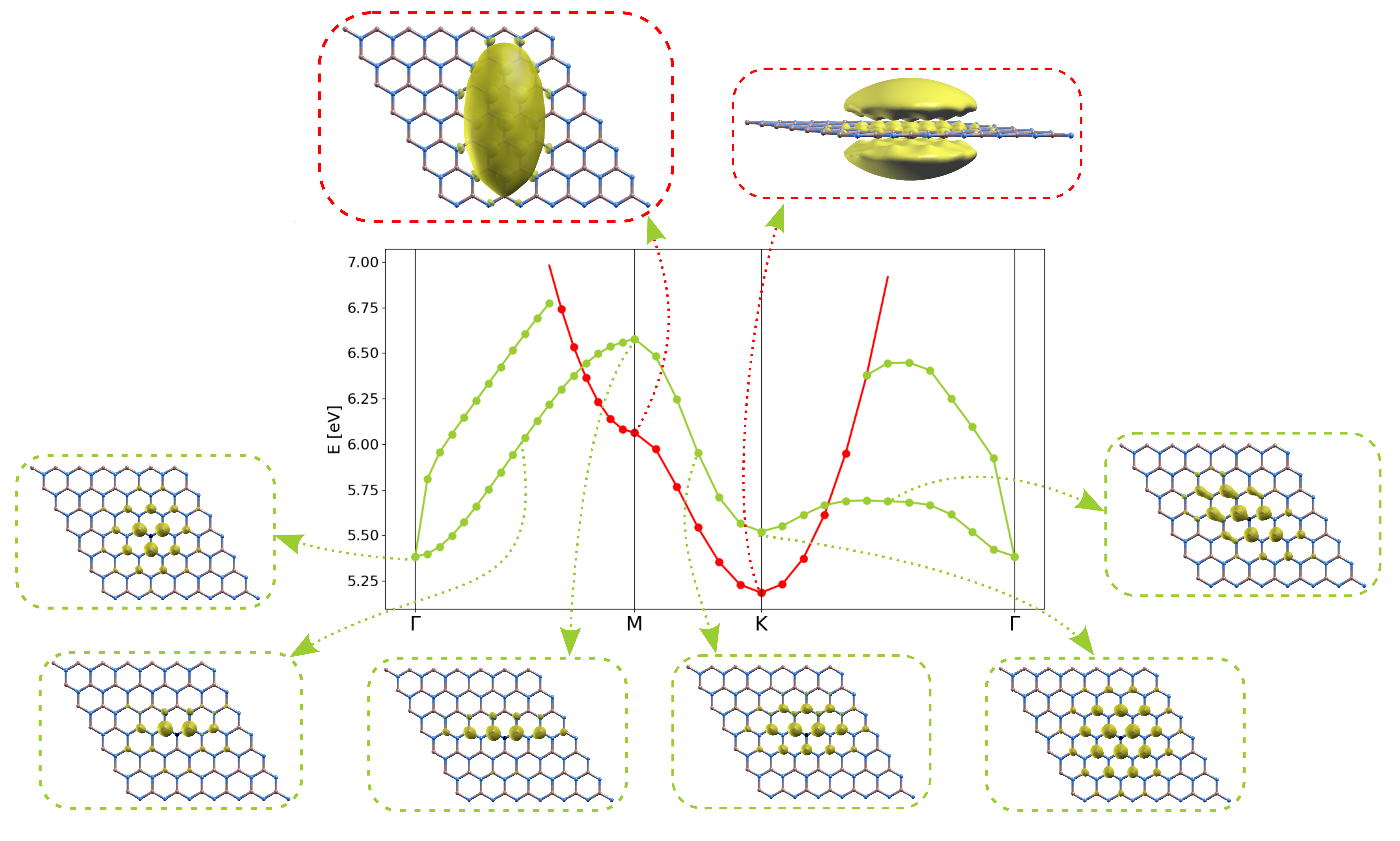}
	\caption{Details of the exciton dispersion of monolayer hexagonal BN. The insets show the spatial localization of the exciton wavefunction at several different $q$-point and branches (this is obtained by fixing the hole position on top of a nitrogen atom, i.e. on a valence $p_z$ orbital, and plotting the resulting electron density). As evidenced in the insets, the red branch in the dispersion plot is due to the nearly-free electron states (involving conduction bands with $\sigma^*$ character), while the green branches originate from the optically active $\pi-\pi^*$ band transitions.\label{fig:exc_wf}}
\end{figure*}

Using the Kohn-Sham band structure as a starting point, we employed MBPT as implemented in the Yambo code\cite{yambo} to calculate quasiparticle band structures within the G$_0$W$_0$ approximation\cite{aryasetiawan1998gw} and the excitonic optical response functions using the Bethe-Salpeter Equation.\cite{strinati}
All the many-body operators that enter in these calculations are expanded in a Kohn-Sham basis set. Therefore, in order to have converged results, we diagonalized the Kohn-Sham Hamiltonian for a large number of bands that were then used to build the electronic Green's function $G$ and the dielectric matrix $\epsilon$. In table~\ref{tab:parms} we also report the cutoff used in the construction of the dielectric matrix. Finally, in order to speed up convergence with respect to the empty states, we used a terminator for both $\epsilon$ and $G$.\cite{bruneval2008accurate}
The BSE was constructed using a static kernel derived from the GW self-energy within the Tamm-Dancoff approximation.\cite{strinati}  We include only the conduction and valence bands close to the gap in such a way to get converged absorption and emission spectra. The BSE was solved for a grid of transferred momenta, the $\QQ$ in Eq.~\eqref{bse_q}, identical to the phonons grid. The phonon and exciton dispersion relations for the different systems are reported in Fig.~\ref{fig:exc_wf} and in the SM.

Luminescence spectra were calculated using Eq.~\eqref{lum_eq}. We first built the exciton-phonon matrix elements using the results obtained from BSE and DFPT, as indicated in the scheme in Fig.~\ref{fig:scheme}. We selected a number of ``initial'' excitons at finite $\QQ$ (indices $\beta$ in Eq.~\eqref{lum_eq}), that scatter with phonons (indices $\mu,\qq$) and end up in the ``final'' excitonic states at $\QQ=0$ (indices $\lambda$). All phonon modes and transferred momenta were included in these calculations. Note that both the the electronic and transferred momenta ($k$-grid and $q$-grid, respectively), were computed on the irreducible parts of the respective Brillouin zones (BZs).
The exciton-phonon coupling matrix elements $\mathcal{G}$ from Eq.~\eqref{eq:Gexcphon} were then expanded in the full BZs by symmetry transformations applied to the electron-phonon matrix elements $g$ and the excitonic coefficients $A$. In this way, we are able to strongly speed up exciton-phonon calculations, which would otherwise require the switching off of all crystal symmetries at the DFPT and MBPT levels.
Then, we interpolated both exciton dispersions -- using a smooth Fourier interpolation\cite{pickett1988smooth} -- and phonon energies -- using the force constants method implemented in Quantum Espresso -- on a finer grid. Subsequently, these interpolated values were used in a double-grid approach to speed up convergence of the luminescence spectra with respect to the transferred momenta grid as explained in App.~\ref{doublegrid}.

We also mention that the two quantities which enter the definition of the exciton-phonon matrix elements, i.e. the electron-phonon matrix elements and the BSE eigenvectors, bring random phases that depend on the fact that different sets of KS wavefunctions were used to generate them. It is a non-trivial technical and numerical point to have these phases consistently accounted for.
Indeed, some DFPT implementations, (like Quantum Espresso), recalculate the KS wavefunctions at $\kk+\qq$ for each $\qq$. Instead, a single set of wavefunctions is used to define the BSE matrix at any momentum $\QQ$ in the Yambo code, where the $\kk+\qq$ wavefunctions are obtained by symmetry transformations, thus imposing a specific choice of the relative phase between the wavefunctions. This difference causes a phase mismatch in the definition of the exciton-phonon matrix elements, Eq.~\eqref{eq:Gexcphon}, because both the electron-phonon matrix element and the excitonic coefficients enter as full complex numbers. This issue remains also if the electron-phonon matrix elements are obtained via Wannier interpolation\cite{chen2020exciton}, since the wavefunction used to construct the excitonic matrix would be different from the ones resulting via the Wannier procedure. In this case the interpolation process should be modified by fixing the wavefunction phases\cite{giustino2007wannier}.
The phase mismatch is not present in calculations based on finite differences\cite{paleari2018excitons,lechifflart2022excitons} because in this case exciton-phonon coupling is directly calculated as a derivative of the BSE matrix elements on a supercell.
However, these types of calculations are restricted to a single $\qq$-vector.
In the case of hBN luminescence, we verified that the phase mismatch only gives small changes in the numerical results (by testing different sets of wavefunctions with different random phases; see also the discussion of the numerical results). For this reason we do not address this issue in the present paper, but we leave it for future works.

\begin{table}
\begin{center}
    \begin{tabular}{ | c | c | c |}
    \hline
	     Parameter/System &  m-hBN & hBN \\ \hline
    \hline
    $\qq$/$\kk$-grid & $36 \times 36 \times 1$ & $18 \times 18 \times 6$  \\ \hline
    $a$ & 2.48 \r{A} & 2.50 \r{A}   \\ \hline
    $c$ & - & 3.25 \r{A}   \\ \hline
	    GW/ $\varepsilon(\omega)$ bands   & 200 & 210 \\ \hline
	    $\varepsilon(\omega)$ cutoff  & 10~Ha & 8~Ha \\ \hline
    BSE bands & 3-6 & 5-12 \\ \hline
    Excitons ($\beta \rightarrow \lambda$) & $8\rightarrow 2$ & $12\rightarrow 4$ \\ \hline
    Double grid & $108 \times 108 \times 1 $ & $54 \times 54 \times 18 $ \\ \hline
    \end{tabular}
	\caption{List of the relevant computational parameters entering the calculation of excitons, phonons and their coupling ($a$ is the planar lattice parameter and $c$ the interlayer distance).\label{tab:parms}}
\end{center}
\end{table}

\begin{figure}
    \centering
    \includegraphics[width=0.48\textwidth]{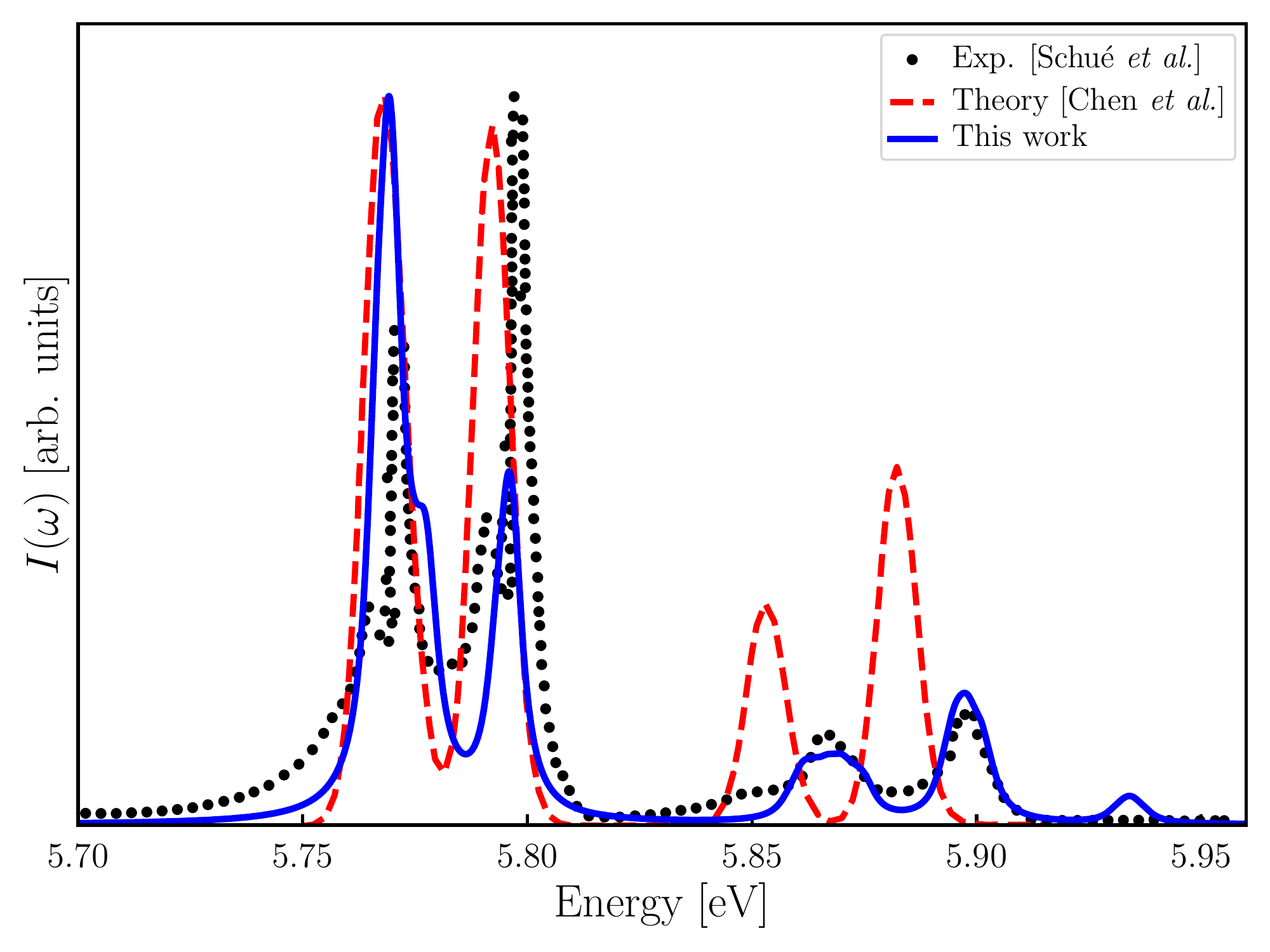}
	\caption{Calculated phonon-assisted luminescence in bulk AA$^\prime$ hBN. The spectrum (blue line) has been shifted to match the lowest emission peak of the experiment (black dots).\cite{schue2018direct} For comparison, we report also the theoretical results of Ref.~\onlinecite{chen2020exciton} (dashed red line). \label{fig:hbn_lum} }
\end{figure}

\section{Results}\label{results}
We now turn to the luminescence results obtained with the theory and implementations presented in the previous section on two distinct cases, bulk and monolayer hBN.

\subsection{Bulk hexagonal boron nitride}
The luminescence of hBN in the AA$^\prime$ stacking has been measured in several experiments\cite{cassabois2016hexagonal,schue2018direct} and subject of many theoretical studies\cite{paleari2018excitons,cannuccia2019theory,lechifflart2022excitons,chen2020exciton}. Therefore it is an excellent case for testing our theory and its approximations. We calculated luminescence using Eq.~\eqref{lum_eq} and all the computational details described in the previous section. Concerning the excitonic effective temperature that enters the luminescence equation, we extrapolated it as a function of the lattice temperature using the data of Ref.~[\onlinecite{cassabois2016hexagonal}] (see sec. V of the SM) and used a lattice temperature of $6$ K as reported in the experiments.

In Fig.~\ref{fig:hbn_lum} we report the results along with the theoretical spectrum of Chen \emph{et al.}\cite{chen2020exciton} and the experimental measurements by Schu\'e \emph{et al.}\cite{schue2018direct}
By comparing experiments with our calculations, we see that we can correctly reproduce the position of the emission peaks, but not exactly their shape and intensity.  Note that our theory does not include multiple phonon scattering, which is responsible for the splitting of the first two peaks and their asymmetric shape.\cite{vuong2017overtones}

As for the intensity of the peaks, however, we note that the lowest-energy two-peak structure, attributed to longitudinal and transverse optical phonon modes (LO and TO; four modes are contributing to this structure), are completely inverted in intensity compared to the experiment. The situation is better for the higher-energy satellites assisted by the longitudinal and transverse acoustic modes (LA and TA along with their almost-degenerate optical companion modes), responsible for the third and four peaks.
The full decomposition of the luminescence in terms of the different phonon modes is presented in the SM.

The previous calculation by Chen \emph{et al.} does not seem to improve much the comparison with the experiments.
Different causes may be at the root of these discrepancies.
First of all, the absolute value of the electron-phonon coupling matrix elements can be underestimated for particular phonon modes due to the use of local exchange correlation functionals.\cite{faber2015exploring}
Second, the above mentioned phase mismatching problem also induces some variability in the intensity ratios.
We performed different tests, which correspond to different choices of the random phases, on this system, e.g. working without crystal symmetries and implementing different conventions for the momentum conservation in Eq.~\eqref{eq:Gexcphon}, and we found no relevant changes in final luminescence spectra of Fig.~\ref{fig:hbn_lum}.
An example of the small errors present in the computed spectra is the small peak due to the ZA/ZO mode, which is the last peak on the right in Fig.~\ref{fig:hbn_lum}. Such phonon-mediated transition should be vanishing by symmetry considerations\cite{paleari2019exciton,cassabois2016hexagonal}. The obtained small intensity could be due to the previously mentioned numerical issues, which induce some small symmetry breaking.

\begin{figure}
    \centering
    \includegraphics[width=0.48\textwidth]{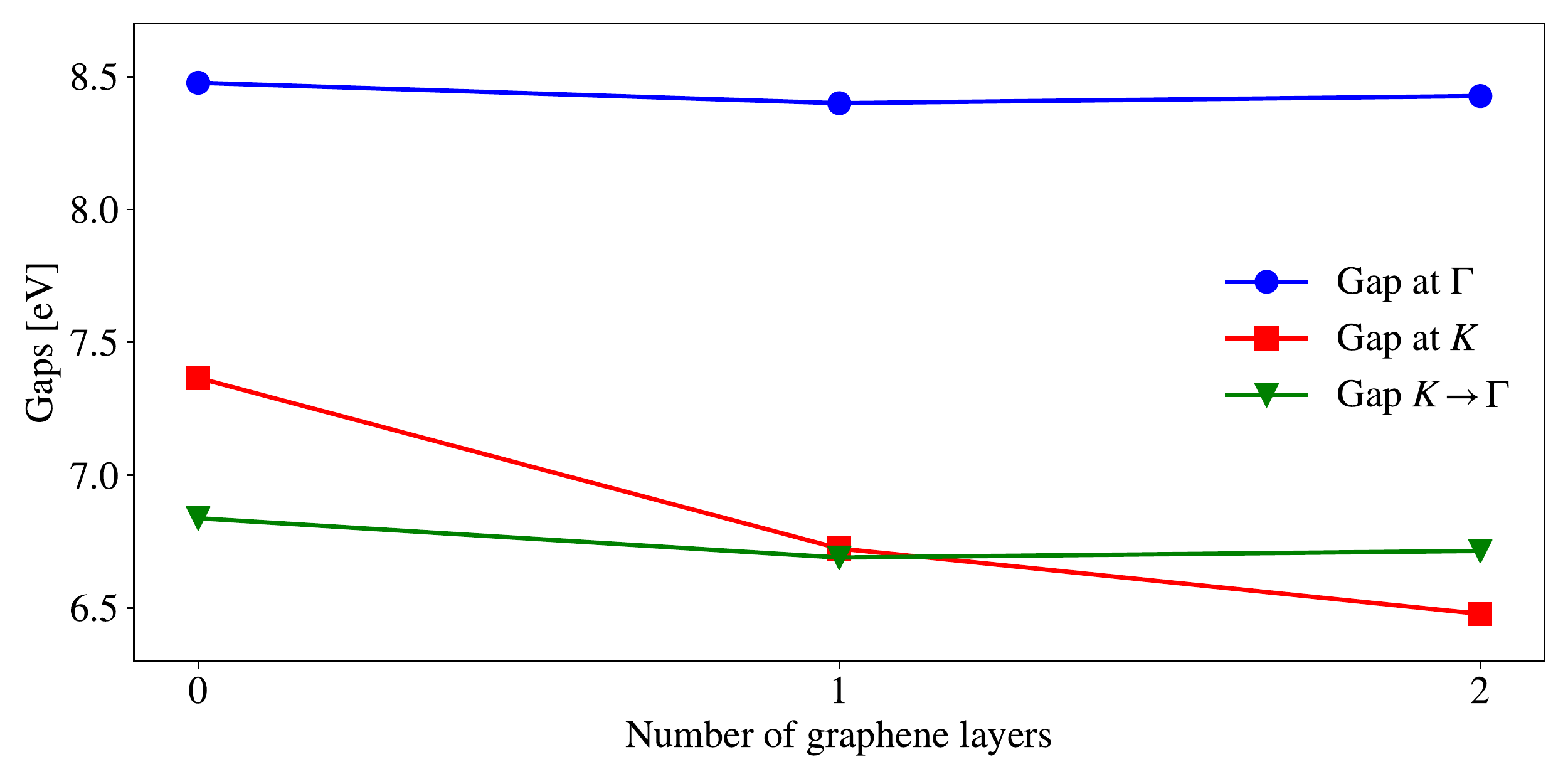}
        \caption{Band gaps of m-hBN as a function of the number of graphene layers\label{gap_vs_layers}. The large direct gap at $\Gamma$ is in blue, the indirect $\pi\rightarrow\sigma^*$, i.e. $K\rightarrow\Gamma$, is in green and the smallest $K\rightarrow K$ direct gap is in red.}
\end{figure}

\subsection{Monolayer \hbn}
The electronic and optical properties of monolayer (m-hBN) have been the subject of numerous studies using both \emph{ab initio} and semi-empirical methods.\cite{galvani2016excitons}
Within DFT, with the LDA exchange-correlation functionals, m-hBN is a direct band gap material at high-symmetry point $K$, but the G$_0$W$_0$ corrections change its gap from direct to indirect, going from $K$ to $\Gamma$.\cite{prete2020giant}
We verified that the system remains indirect even at the semi-self-consistent ``eigenvalue-GW'' level, see Sec. I of the SM.
This indirect gap is due to the presence of nearly-free electron-like states at $\Gamma$. In fact, for these $\sigma^*$-like states, the screened Hartree-Fock correction provided by the $GW$ self-energy, which opens the band gap, is much smaller compared to that of the $\pi$-like states around $K$ and $M$.  The nearly-free electron states have been investigated in BN nanotubes and m-hBN\cite{blase1994stability,Blase1995monolayer}, but only at the independent-particles, DFT level. They may provide a possible mechanism for luminescence quenching.

Despite the presence of these states at $\Gamma$, the optical properties of BN-based systems are actually dictated by the $\pi$ bands around $K$ and $M$, and this remains true for m-hBN.
The optical spectrum of m-hBN is characterized by a strong doubly degenerate excitonic peak of symmetry $E$ at about $6$~eV. Exciton dispersions have been reported in several papers\cite{koskelo2017excitons,sponza2018direct}. In Fig.~\ref{fig:exc_wf} we also report our calculated dispersion along selected high-symmetry lines, starting from the quasiparticle band structure.

\begin{figure*}[ht]
    \centering
    \includegraphics[width=0.85\textwidth]{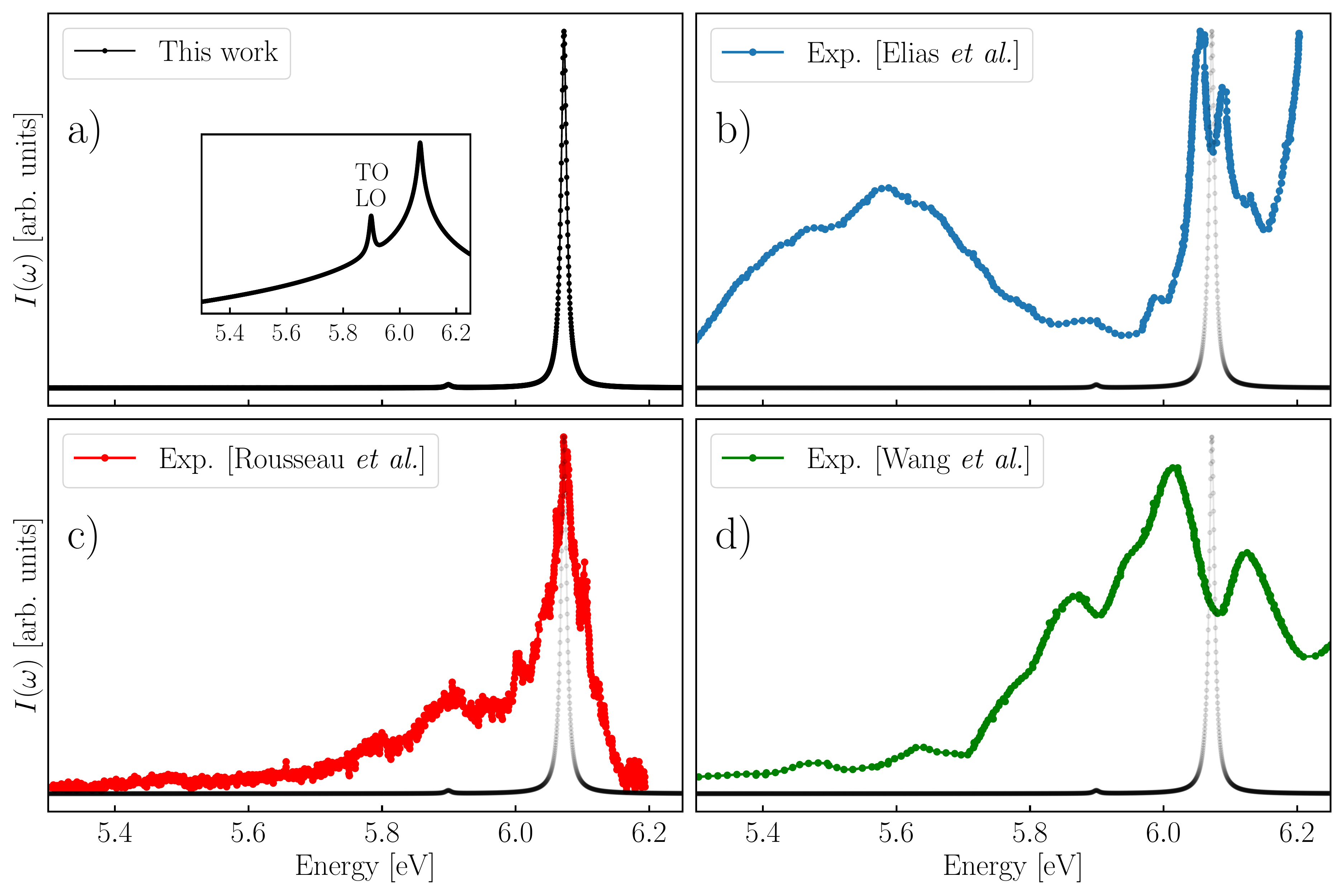}
	\caption{Calculated luminescence spectrum of monolayer hBN (a) compared to the experimental results of Ref.~\onlinecite{elias2019direct}(b), Ref.~\onlinecite{rousseau2021monolayer}(c) and Ref.~\onlinecite{wang2022scalable}(d). The theoretical spectrum has been shifted to match the main experimental peak (c). For clarity, we have plotted the theoretical spectrum next to each experimental result. In the inset of panel (a) we show the theoretical spectrum in a logarithmic scale, revealing the presence of a small phonon replica.\label{fig:1l_lum}}
\end{figure*}

We found that excitons at momentum $\qq=K$ have a lower energy than the direct exciton at $\qq=0$, a feature inherited from the indirectness of the quasiparticle structure. In fact, these low-energy excitons are due to transitions towards the nearly-free electron states at $\Gamma$.
These new excitonic states are clearly distinguishable from the ``standard'' BN excitons by plotting their wavefunctions in real space, as it is done in Fig.~\ref{fig:exc_wf} for several different center-of-mass momenta of the various states. While the usual $\pi \rightarrow \pi^*$-derived states (green exciton ``bands'' in the figure) display the electronic density strongly localized on the boron sublattice, when the hole is fixed on top of a nitrogen, the $\pi \rightarrow \sigma^*$-derived states (red exciton ``band'') present an electron density strongly delocalised away from the layer plane. This is a clear signature of nearly-free electron character.

In order to see if we can really expect an optical experimental signature from these states -- that will make the system functionally indirect -- we decided to investigate how the presence of a substrate modifies their position with respect to the direct gap, compared to the freestanding m-hBN. We included a graphitic substrate in the simulation, analogous to the one used in some of the experiments, and found that it lowers the direct gap at $K$ much more than the indirect one, actually making the system a true direct band gap insulator again. This is most likely due to the stronger interaction of the $p_z$ orbitals of boron with those of graphene while the planar $\sigma^*$ states are less affected. The dependence of the m-BN gaps on the number of substrate graphene layers is shown in Fig.~\ref{gap_vs_layers} (calculation details are given in the SM).
Therefore, we expect that these states at $\Gamma$ will not contribute to the luminescence in a realistic experiment where m-hBN is deposited on a substrate.
In order to simulate luminescence from an ideal m-hBN deposited on a substrate we started from the LDA band structure and applied a scissor operator that allows us to maintain the direct nature of m-hBN (i.e., removing the red ``band'' in Fig. \ref{fig:exc_wf}, see also Sec.~III of the SI). Then we calculated the luminescence spectra using a lattice temperature of $10$ K, while the effective excitonic temperature was estimated from the lattice one using the relation in Sec.~IV of the SM.

In panel $(a)$ of Fig.~\ref{fig:1l_lum} we report our luminescence calculations of a single layer m-hBN compared to the measurements of Refs.~\onlinecite{elias2019direct,wang2022scalable,rousseau2021monolayer}, panels $(b),(c),(d)$. Beside the main direct emission peak, we found a zero-momentum phonon replica due to the TO and LO phonons (see inset in logarithmic scale in the panel $(a)$).
However, the satellite intensity is very low and barely visible on a linear scale.
Therefore, it cannot explain the additional peaks seen in some of the experiments. In our calculation we also included possible indirect transition from excitons with momentum corresponding to the $K$ point due to the $\pi \rightarrow \pi^*$ transitions (relative minimum of the green curve in Fig.~\ref{fig:exc_wf}). However, we found that due to the large energy difference, $0.14$~eV, between direct and indirect excitons, the contribution of these latter states to the luminescence spectrum is null.\footnote{We did not consider polaritonic effects that could slightly modify the luminescence spectra, see Ref.~\onlinecite{henriques2019absorption} for a discussion.}

As a sanity check, we considered the possibility that the distance between the exciton $\KK$ and $\Gamma$ is not well reproduced by our calculations and analysed the matrix elements of the phonon-assisted transitions between $\KK$ and $\Gamma$. We found that these are in any case too small to explain the additional peak seen close to the main one in the graphite experiments.


In the light of these results, let us now discuss the different experiments. The details of the three luminescence spectra reported in the literature, see Fig.~\ref{fig:1l_lum}(b-d), are the following: two of them feature m-hBN deposited on graphite,\cite{elias2019direct,wang2022scalable} and one on silicon oxide.\cite{rousseau2021monolayer}
With the graphite substrate, multiple peaks are visible.
These peaks have been attributed to various causes, which we will briefly summarize here. In the work of Elias \emph{et al.},\cite{elias2019direct} the possibility of additional replicas due to indirect excitons at $K$ was considered.
In the work of Rousseau \emph{et al.}\cite{rousseau2021monolayer} they put forward the possible presence of bubbles in the sample as cause of the additional peaks. 
Finally, in the article of Wang \emph{et al.}\cite{wang2022scalable}, these additional peaks were attributed to the presence of multilayer BN regions and/or defects.

Our theoretical work allows us to rule out the first hypothesis since, as shown above, the energy difference between $\Gamma$ and $\KK$ is large and the phonon-assisted transitions have too low an intensity to have indirect excitons visible in luminescence in the energy range where the second experimental peak appears, while the $\pi\rightarrow\sigma^*$ transitions seem to play no role. Regarding the effect of bubbles on the luminescence spectra, recently some of us have shown that strain can induce shifts of the luminescence spectra.\cite{lechifflart2022excitons}
Yet, in order to obtain a significant effect, the strain must be very large, and in addition it is difficult to explain with strain the presence of two well-defined peaks, like those visible in the spectra. 
Finally, there is the hypothesis of defects or multi-layers BN. We think this is the most plausible hypothesis, because it has been shown that some defects can produce levels close to the main exciton\cite{attaccalite2011coupling}, and multi-layers BN induce splittings of the main peak\cite{paleari2018excitons}. 
Finally, note that the presence of defects or edges, which break translational symmetry, could make the indirect exciton visible without the need for phonon mediation.\cite{feierabend2017proposal}

\section{Conclusions} \label{conclusions}
In this manuscript, we presented a first-principles methodology to calculate phonon-assisted luminescence in exciton-dominated materials. It is based on a dynamical correction to the static Bethe-Salpeter equation given by an excitonic self-energy term describing exciton-phonon interaction. Using this self-energy, we obtained a formula for the optical response that contains corrections up to first order in the exciton-phonon coupling.
Unlike previous formulations, we are also able to calculate the renormalisation factor for direct transitions, which allows for a quantitative comparison between direct and phonon-assisted emission signatures.
From the optical response function, and employing a steady-state approximation, we obtained a formula for the phonon-assisted luminescence. All ingredients that enter in this formulation have been calculated \emph{ab initio}.
We first validate our approach on bulk \hbn, where clear and well-established experiments exist. We then applied this approach to the BN single layer, where recent discordant photoluminescence measurements were reported independently by different groups. In m-hBN we found that the luminescence spectrum is dominated by the single direct peak only and phonon replicas, while present, have negligible intensity. In addition, 
phonon-assisted transitions from the lowest indirect exciton remain too low in intensity to explain the measured spectra. Therefore, we rule out phonon-assisted processes as the cause of the additional spectral fine structure sometimes seen in experiment.
We support the interpretation that this fine structure is not intrinsic, nor due only to substrate effects, but depends on sample quality.
Finally, we mention that our methodology based on the dynamical self-energy is fully implemented in the \texttt{Yambo} code and applicable to other systems of interest. This formulation allows one to obtain more observables than just the luminescence presented here, such as phonon-assisted absorption and exciton linewidths and relaxation rates.
Thus, we hope that this work will motivate both further experimental measurements on BN-based systems and theoretical advancements in the efficient modeling of materials with strong exciton-phonon coupling.

\section*{Acknowledgments}
The research leading to these results has  received funding from the European Union Seventh Framework Program under grant agreement no. 785219 Graphene Core2. This publication is based upon work from COST Action TUMIEE CA17126, supported by COST (European Cooperation in Science and Technology). F.P. acknowledges the European Union project: MaX {\em Materials design at the eXascale} H2020-INFRAEDI-2018-1, grant agreement n. 824143. This work has been performed under the Project HPC-EUROPA3 (INFRAIA-2016-1-730897), with the support of the EC Research Innovation Action under the H2020 Programme; the authors gratefully acknowledges the support of CNR-Nano group in Modena, in particular D. Varsano and the computer resources and technical support provided by CINECA. The authors acknowledge A. Saul and K. Boukari for the management of the computer cluster \emph{Rosa}, and F. Ducastelle, M. Zanfrognini, E. Cannuccia, J. Barjon and A. Loiseau for useful discussions. CA thanks L. Wirtz for comments and remarks.
\\

\appendix
\section{Double-grid for the exciton-phonon problem}\label{doublegrid}
Convergence of luminescence spectra can require a fine sampling of the $\qq$ space. For this reason we interpolated phonon and exciton energies on a finer $\tilde q$-grid with respect to the coupling matrix element calculations.
Then these finer grids are used to average out the denominators appearing in the luminescence formula as:
\bean
\frac{1}{W^{\pm}_{\lambda, \beta,  \qq, \mu }} =\frac{1}{N_{\tilde q}} \sum_{\tilde q \in \qq }\frac{1/2 \pm 1/2 + n_{\tilde q,\mu} }{ \omega - (E_{\tilde q,\beta} \mp \omega_{\tilde q \mu}) + i\eta } e^{-\frac{E_{\qq,\beta}-E_{min}}{kT_{exc}}}
\eean
using the above defined average denominators we can write the luminescence as
\begin{widetext}
\bean
	{I}^{PL}(\omega)&=& \Im \sum_\lambda \frac{| T_\lambda  |^2}{\pi^2 \hbar c^3} \left \{ \omega^3 n_r(\omega) \frac{1 - R_\lambda}{\omega - E_\lambda + i\eta} e^{-\frac{E_{\lambda}-E_{min}}{kT_{exc}}}+ \sum_{\mu\beta\qq} \omega(\omega \mp 2 \omega_{\qq \mu})^2 n_r(\omega) \left|\mathcal{D}^{\pm}_{\beta\lambda,\qq \mu} \right| \frac{1}{W^{\pm}_{\lambda, \beta,  \qq, \mu }} \right \},
\eean
\end{widetext}
implemented as a sum over the fine-grid $\tilde q$-points in the neighborhood of each coarse-grid $q$-point.

\bibliography{phassisted}
\nolinenumbers
\end{document}


\title{Supplemental Material of: First-principles study of luminescence in hexagonal boron nitride single layer: exciton-phonon coupling and the role of substrate}

\author{Pierre Lechifflart}
\affiliation{\cinam}
\author{Fulvio Paleari}
\affiliation{\cnrmodena}
\author{Davide Sangalli}
\affiliation{\cnr}
\author{Claudio Attaccalite}
\affiliation{\cinam}
\affiliation{\etsf}

\maketitle

In this Supplementary Material we report our computational results for quasiparticle band structures and phonon and exciton dispersion relations. We also comment on the effective excitonic temperature estimation and the exciton occupation factor, as well as showing the contributions of individual phonon modes to the luminescence spectra.

\begin{figure}
    \centering
    \includegraphics[width=0.95\textwidth]{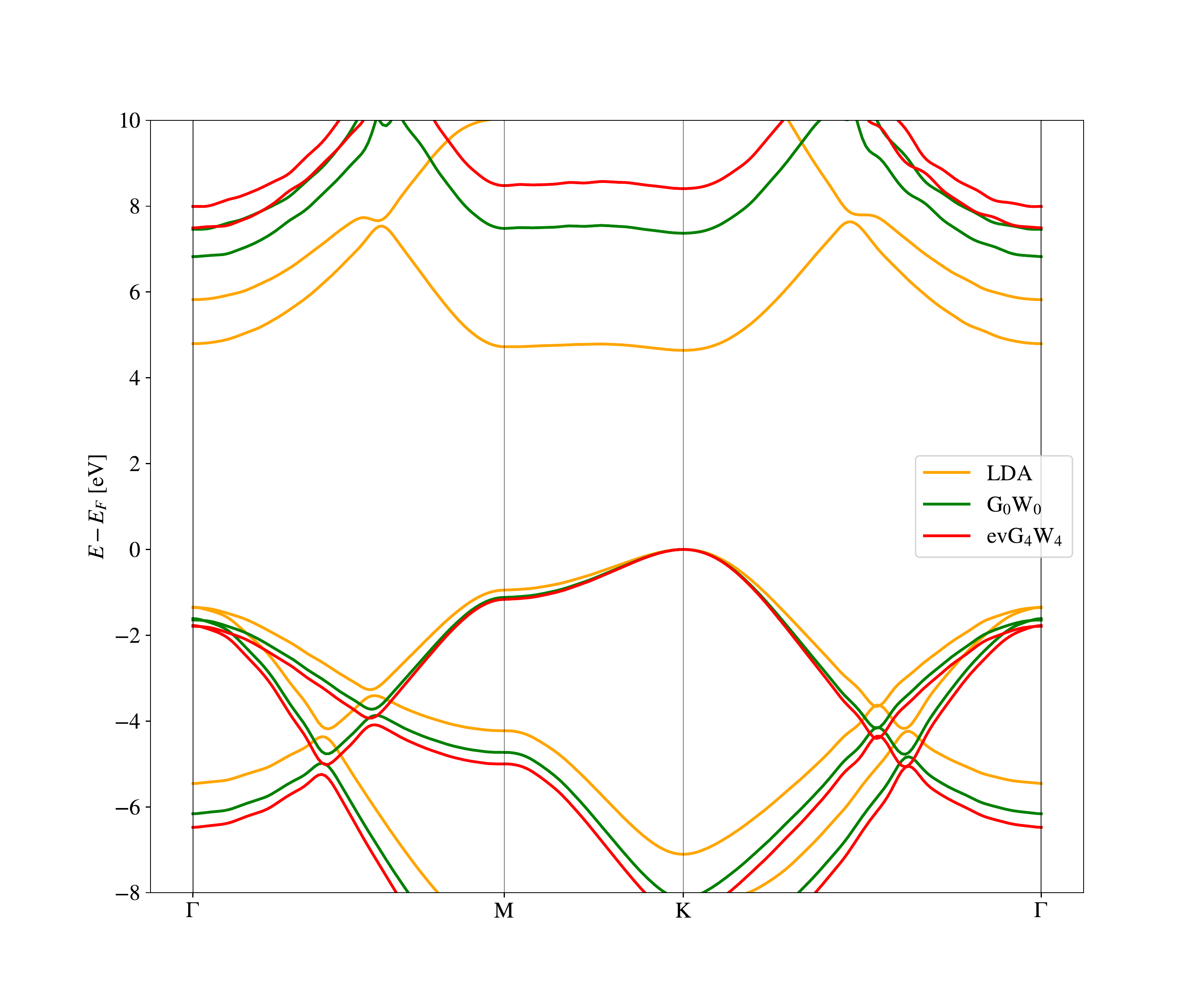}
	\caption{Electronic bands of freestanding monolayer hBN, in DFT (orange), G$_0$W$_0$ (green) and evGW (red).
	} \label{fig:1l_bands_scGW}
\end{figure}

\section{Band structure of monolayer h-BN}

Monolayer hBN is usually regarded as a direct band gap material with the gap localized at the K point.
When calculating the electronic band structure in DFT, for example within the LDA exchange functional, the gap is in fact predicted to be direct at point K. However, when calculating the quasiparticle corrections, the gap becomes indirect between K and the nearly-free electron states at $\Gamma$, see Fig.~\ref{fig:1l_bands_scGW}.  
We report the GW band structure both in the one-shot G$_0$W$_0$ case and with self-consistency applied on eigenvalues only (evGW).\cite{faber2014excited} We found that the nature of the gap does not change with respect to the G$_0$W$_0$ approximation. We also verified that increasing the vacuum separation between the periodic replica does not affect the band orderings.
This result is in agreement with other works that reported an indirect nature for m-hBN\cite{prete2020giant,paleari2019first}. The orbitals corresponding to the nearly-free electron bands at $\Gamma$, that are responsible for the indirect nature of m-hBN, are specially localized above or below the layer plane.\cite{blase1994stability,Blase1995monolayer}.
These nearly-free electron states, being located out of the plane, have practically zero optical dipoles and very small coupling with phonons. This means that they do not contribute to the linear optical properties, although in principle they could be responsible for the quenching of luminescence in particular configurations\cite{schue2016dimensionality}. \\
Calculations of m-hBN on the graphitic substrate have been performed using a $36\times 36 \times 1$ $\kk$-grid, plus a shifted fine double-$\kk$-grid $90 \times 90 \times 1$ for the screened interaction $W$ as described in Ref.~[\onlinecite{kammerlander2012speeding}] and the \emph{slab} cutoff in the z-direction, 7~Ha cutoff for the diectric constant, and conduction bands up to 120~eV. 
\section{Phonon dispersions}
\subsection{Monolayer hBN}
\begin{figure}
    \centering
    \includegraphics[width=0.95\textwidth]{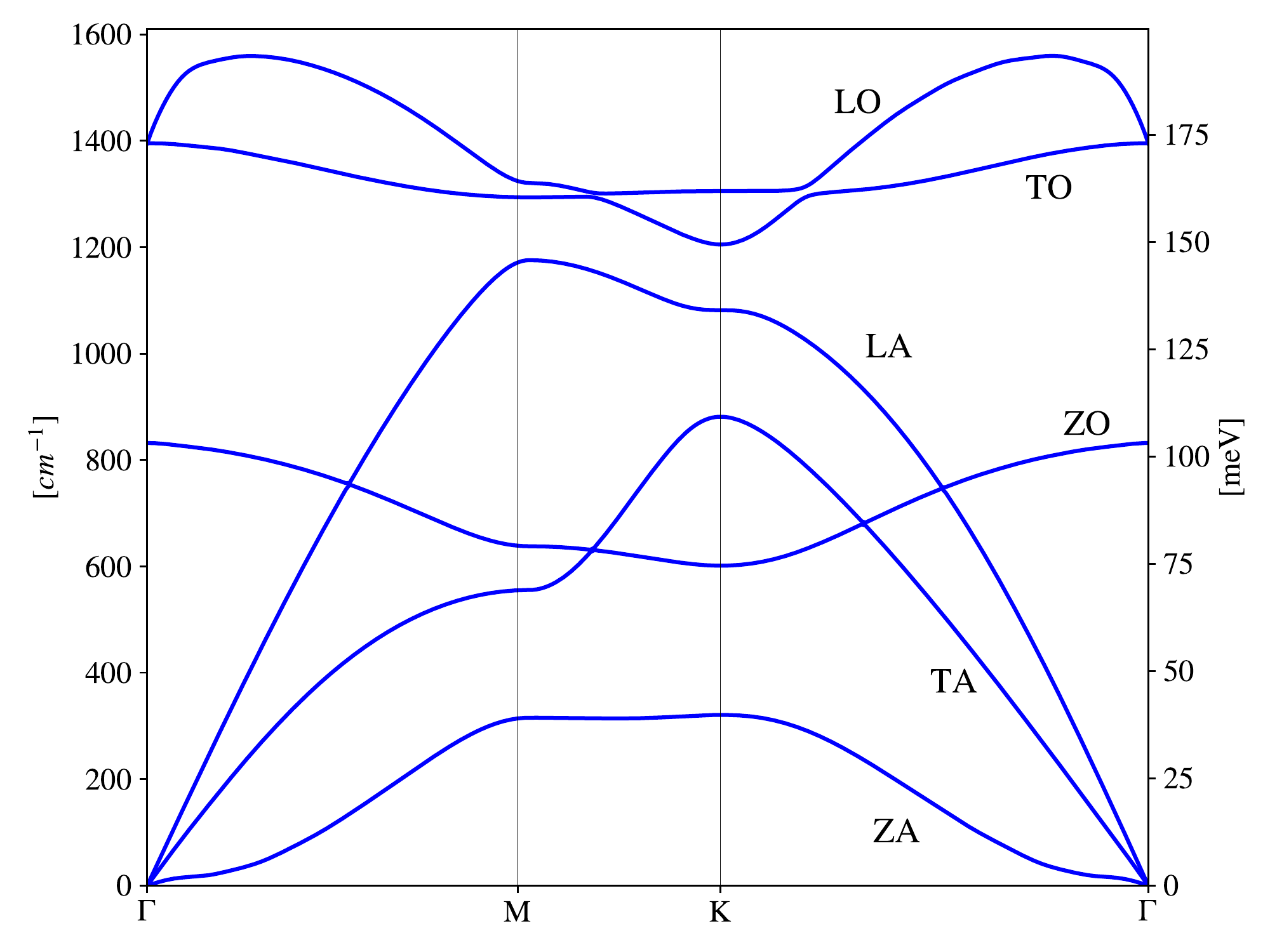}
	\caption{Calculated phonon dispersion for monolayer hBN \label{fig:monolayer_phdisp}}
\end{figure}

Phonons were calculated with a $24 \times 24 \times 1$ grid of transferred momenta with Density Functional Perturbation Theory (DFPT) as implemented in Quantum ESPRESSO. We Fourier-transform the interatomic force constants to real space, adding the non-analytic term and the Coulomb cutoff to correctly treat the longitudinal optical modes in a two-dimensional system. We finally Fourier-transform back to reciprocal space along a high-symmetry path in the Brillouin zone. The resulting phonon dispersion is displayed in Fig. \ref{fig:monolayer_phdisp} and results are in agreement with previous publications.\cite{sohier2017breakdown,wirtz2003ab}

\subsection{Bulk hBN}
\begin{figure}
    \centering
    \includegraphics[width=0.95\textwidth]{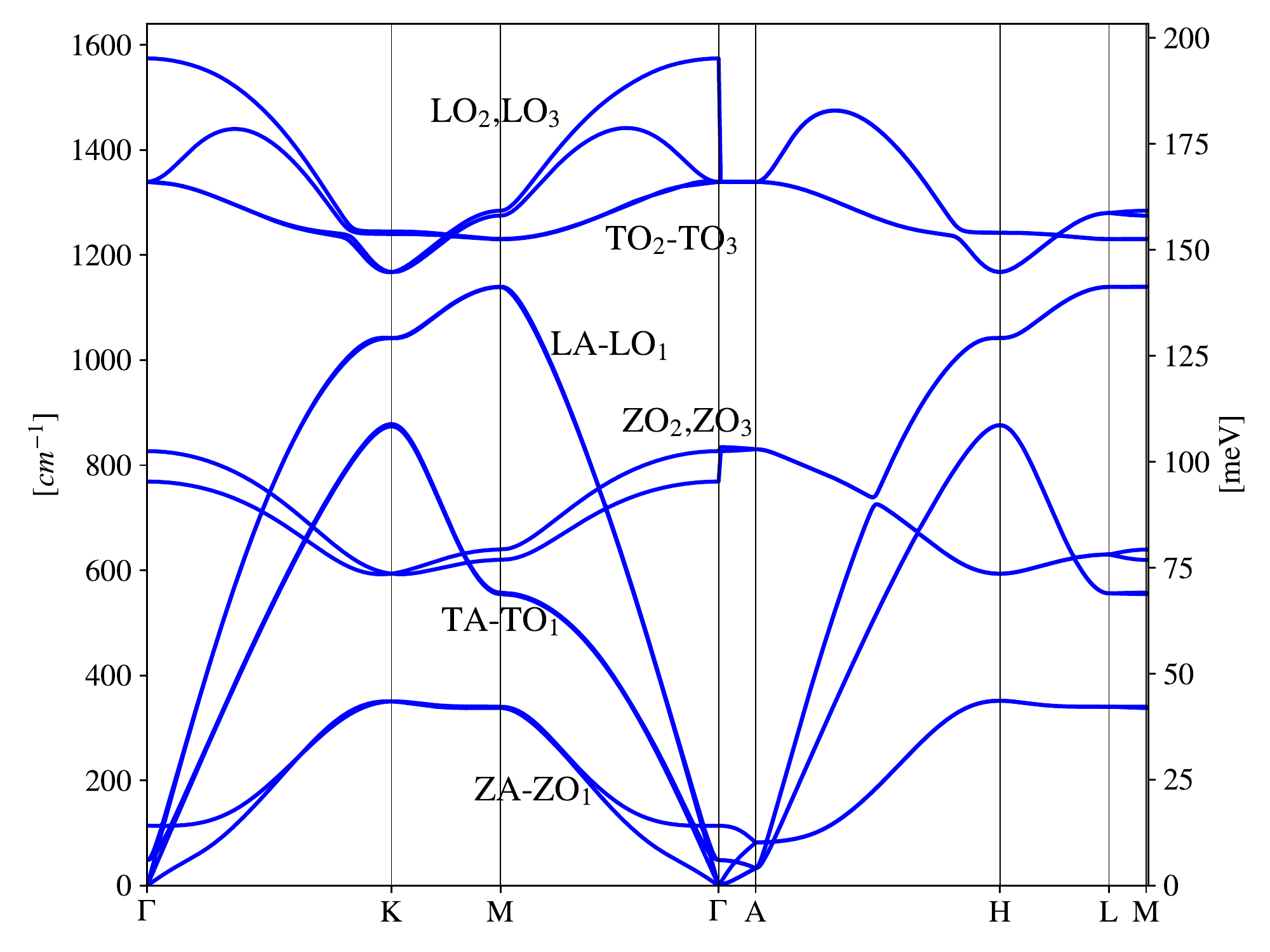}
	\caption{Calculated phonon dispersion for bulk hBN. \label{fig:hbn_phdisp}}
\end{figure}

In this case, phonons were calculated with DFPT using a $12 \times 12 \times 4$ grid of transferred momenta, and the resulting phonon dispersion is displayed in Fig.~\ref{fig:hbn_phdisp}.

\section{Exciton dispersions}
\subsection{Monolayer hBN}
\begin{figure}
    \centering
    \includegraphics[width=0.95\textwidth]{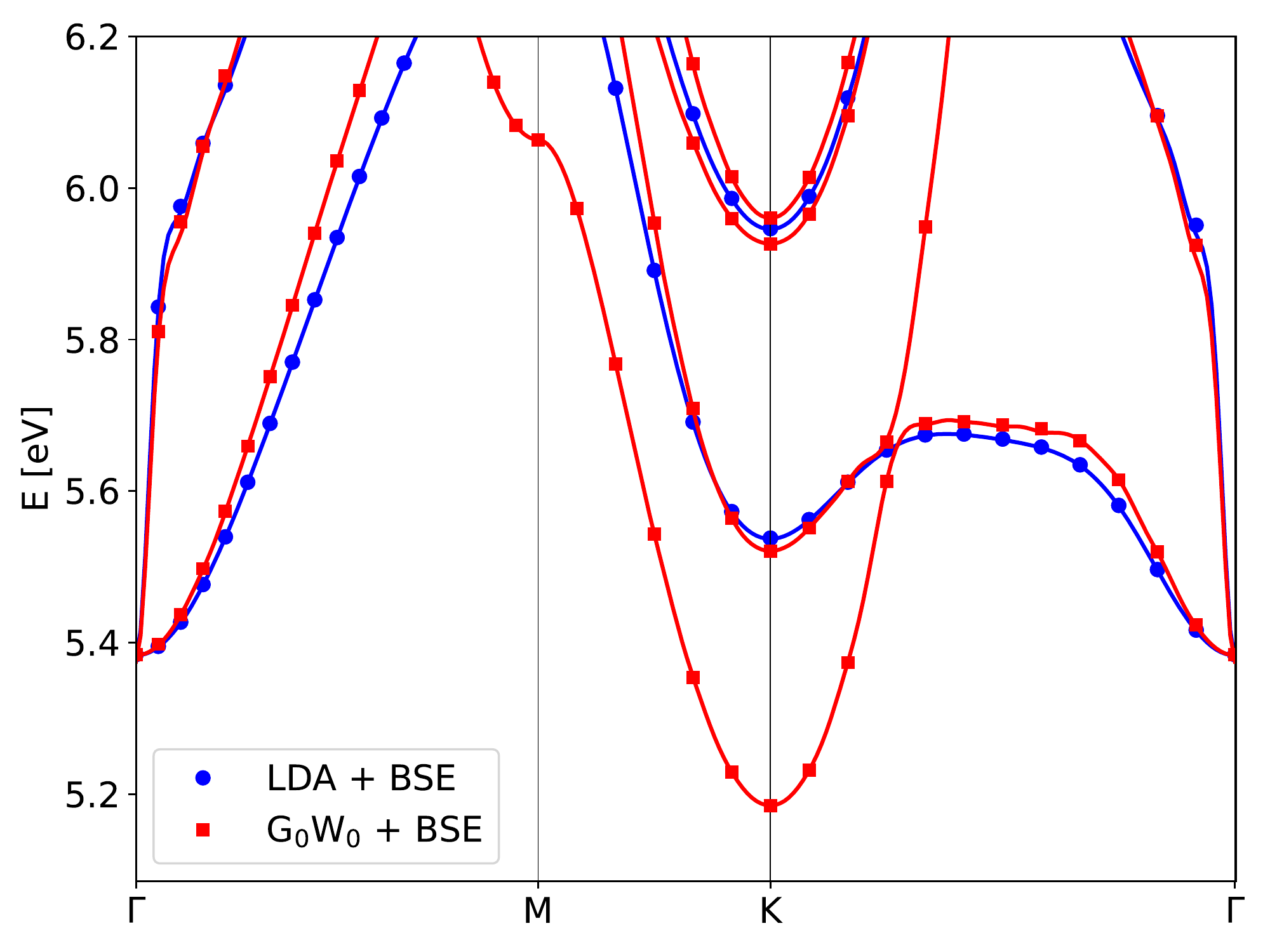}
	\caption{Calculated exciton dispersion for monolayer hBN, starting from either the DFT-LDA eigenvalues (blue) or the G$_0$W$_0$ quasiparticle energies (red). Dots represent the actual BSE data, lines are Fourier interpolations. \label{fig:1l_excdisp_lda_gw}}
\end{figure}

As explained above, the electronic band structure of monolayer hBN is different whether we compute it in DFT or within the G$_0$W$_0$ approximation. Hence, the exciton dispersion will inherit analogous differences whether we use the Kohn-Sham eigenvalues or the quasiparticle energies as a starting point of the Bethe-Salpeter equation, as can be seen in Fig. \ref{fig:1l_excdisp_lda_gw}.

The exciton dispersion computed on top of G$_0$W$_0$ quasiparticle energies exhibits a minimum at K. Upon further investigation of the excitonic wavefunction, see main text, we see that this minimum comes from transitions to nearly-free electron states at $k=\Gamma$ (having parabolic dispersion). Indeed, the wavefunction corresponding to the $q=K$ minimum shows that the electron is delocalized a few angstroms above and below the layer plane. This happens also in the case of bilayer and few-layer hBN systems.\cite{paleari2019first}
On the contrary, the wavefunction of the second-lowest exciton state shows the typical distribution of BN excitons, with the electron mostly localized on boron atoms.\cite{galvani2016excitons}
The nearly-free type excitons have zero oscillator strength at $q=0$ and almost no coupling with phonon modes, thus they do not play a role in optical absorption. As we show in the main text, they are also pushed up in energy from the interaction with the substrate, and therefore they do not influence optical emission either. In order to mimic the substrate effect, we started from the LDA band structure and applied a rigid shift to reproduce the excitonic dispersion without the nearly-free excitons (blue curve in Fig.~\ref{fig:1l_excdisp_lda_gw}). This excitonic dispersion is used in the main text to calculate luminescence.

\subsection{Bulk hBN}
\begin{figure}
    \centering
    \includegraphics[width=0.95\textwidth]{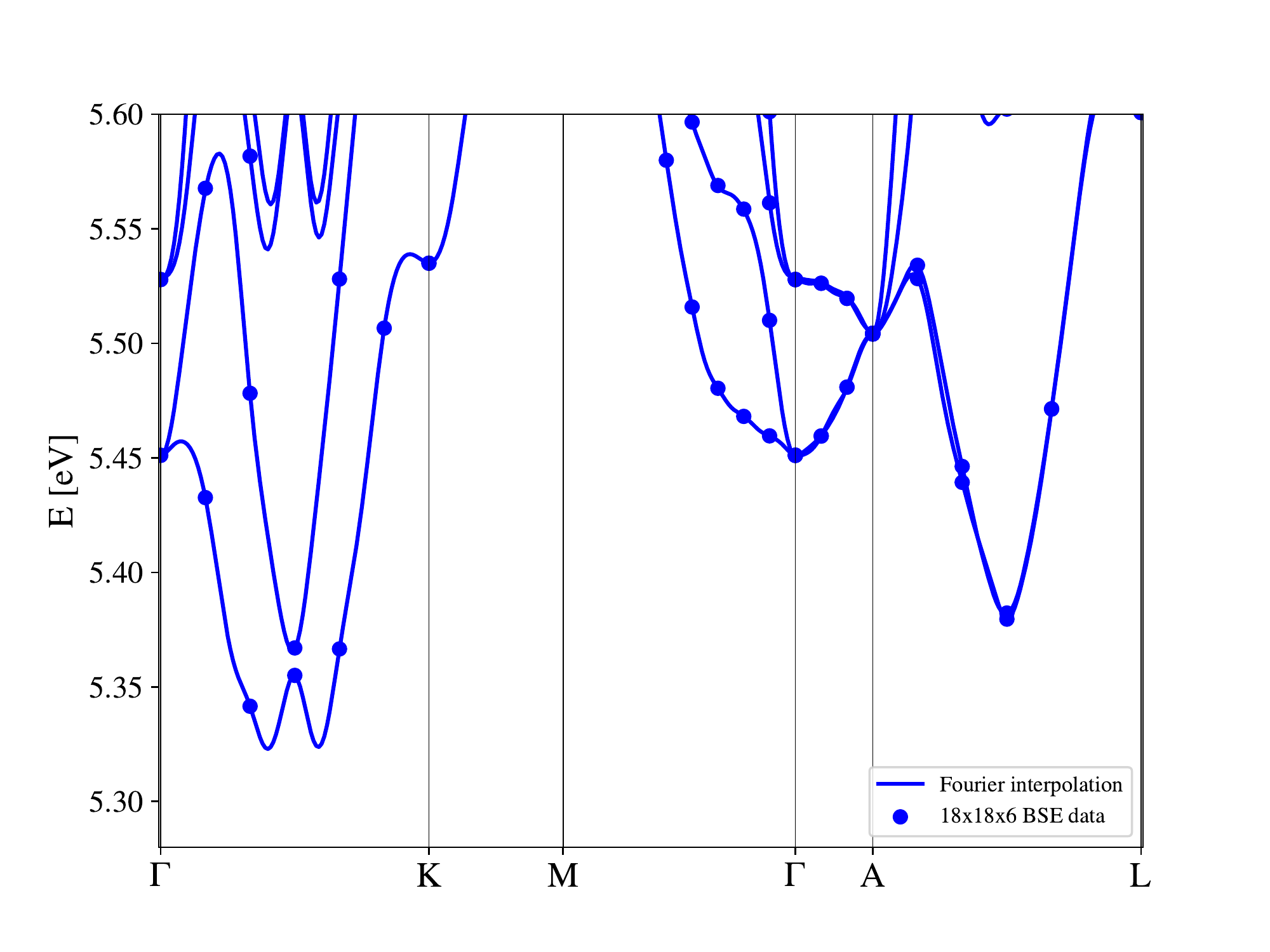}
	\caption{Calculated exciton dispersion for bulk hBN. \label{fig:hbn_excdisp}}
\end{figure}

The BSE at finite $q$ is computed on top of quasiparticle energies at the G$_0$W$_0$ level.
We interpolate the exciton energies with a smooth Fourier method.\cite{pickett1988smooth}
The dispersion displayed in Fig.~\ref{fig:hbn_excdisp} exhibits two minimal exciton energies between $\Gamma$ and K.\cite{sponza2018exciton} These are the excitons that will be populated following the Boltzmann occupation function in our steady-state approximation for luminescence (based on the van Roosbroeck-Shockley relation) in the main text. 
We point out that this minima structure is distributed in a roughly doughnut shape also outside of the strict $\Gamma$K direction. The exact location of the absolute minimum within the doughnut shape is a matter of meV and may fluctuate due to numerical details of the calculation (see points R$_1$ and R$_2$ in the last Section of this SM). The $q$-distribution of excitons in the full BZ is captured by our luminescence calculations, allowing us to properly account for this minima structure, at variance with finite-difference approaches. 
Note that the the two degenerate lowest-lying states at the $\Gamma$ point ($5.45$ eV) are optically forbidden due to inversion symmetry. The optically allowed bright exciton, also doubly degenerate, sit slightly higher in energy at $5.53$ eV.
The lowest-lying excitons along the $\Gamma$-K direction are scattered by phonons into the bright excitons at $\Gamma$ and then recombine, giving rise to indirect peaks in the photo-luminescence spectra.

\section{Excitonic temperature}
\begin{figure}
\centering
\includegraphics[height=0.5\textheight]{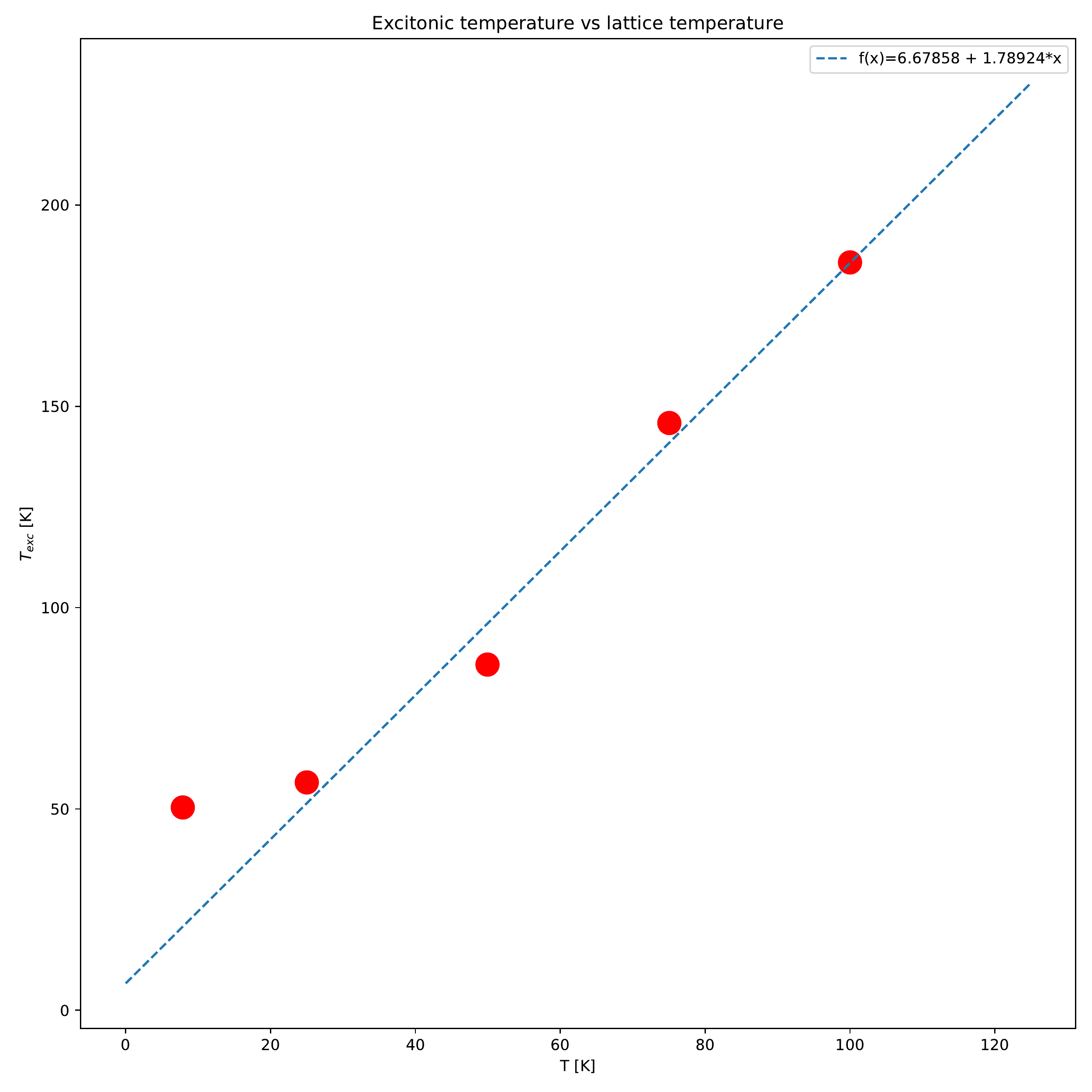}
	\caption{\footnotesize{[Color online] Fit of the excitonic temperature vs lattice temperature from the Cassabois \emph{et al. } data.\cite{cassabois2016hexagonal}.\label{tmpexc}}}
\end{figure}

In order to estimate the effective excitonic temperature we use the data of Cassabois \emph{et al.}~\onlinecite{cassabois2016hexagonal} and fit their results with a linear function excluding the first point. The resulting fit function $T_{exc}=6.68+1.79 T_{latt}$ is shown in Fig.~\ref{tmpexc}. This function is used in the main text to estimate the excitonic temperature from the lattice one for all the different structures.

\section{Photo-luminescence with separated phonon contributions}
We can plot the luminescence intensity by considering individual terms in the sum over phonon modes from Eq.~(14) in the main text. In this way we can determine which phonon mode is responsible for the different satellite peaks in the spectra. 
In the case of hBN, we obtain similar results to the literature, although more peaks are visible due to the presence of more minima in the excitonic dispersion.\cite{paleari2019exciton}

\begin{figure}
\centering
\includegraphics[height=0.5\textheight]{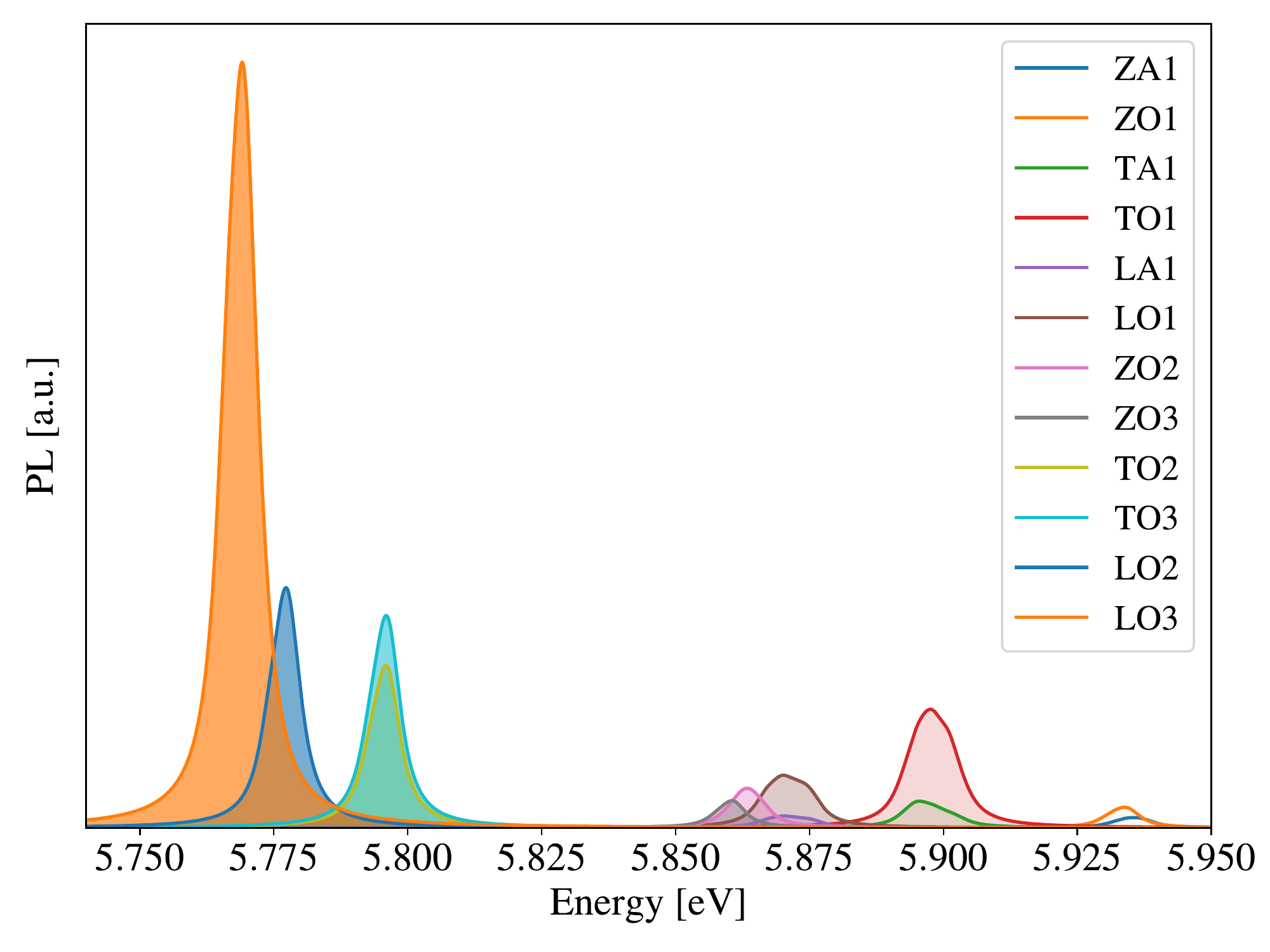}
	\caption{\footnotesize{[Color online] Photoluminescence spectrum of bulk hBN resolved with respect to the different phonon modes, see colors in legend. The labels correspond to those appearing in Fig.~\ref{fig:hbn_phdisp}.\label{pl_hbn_split_phonons}}}
\end{figure}

\section{Excitonic occupation factors} \label{exc_occup}
\begin{figure}
\centering
\includegraphics[height=0.5\textheight]{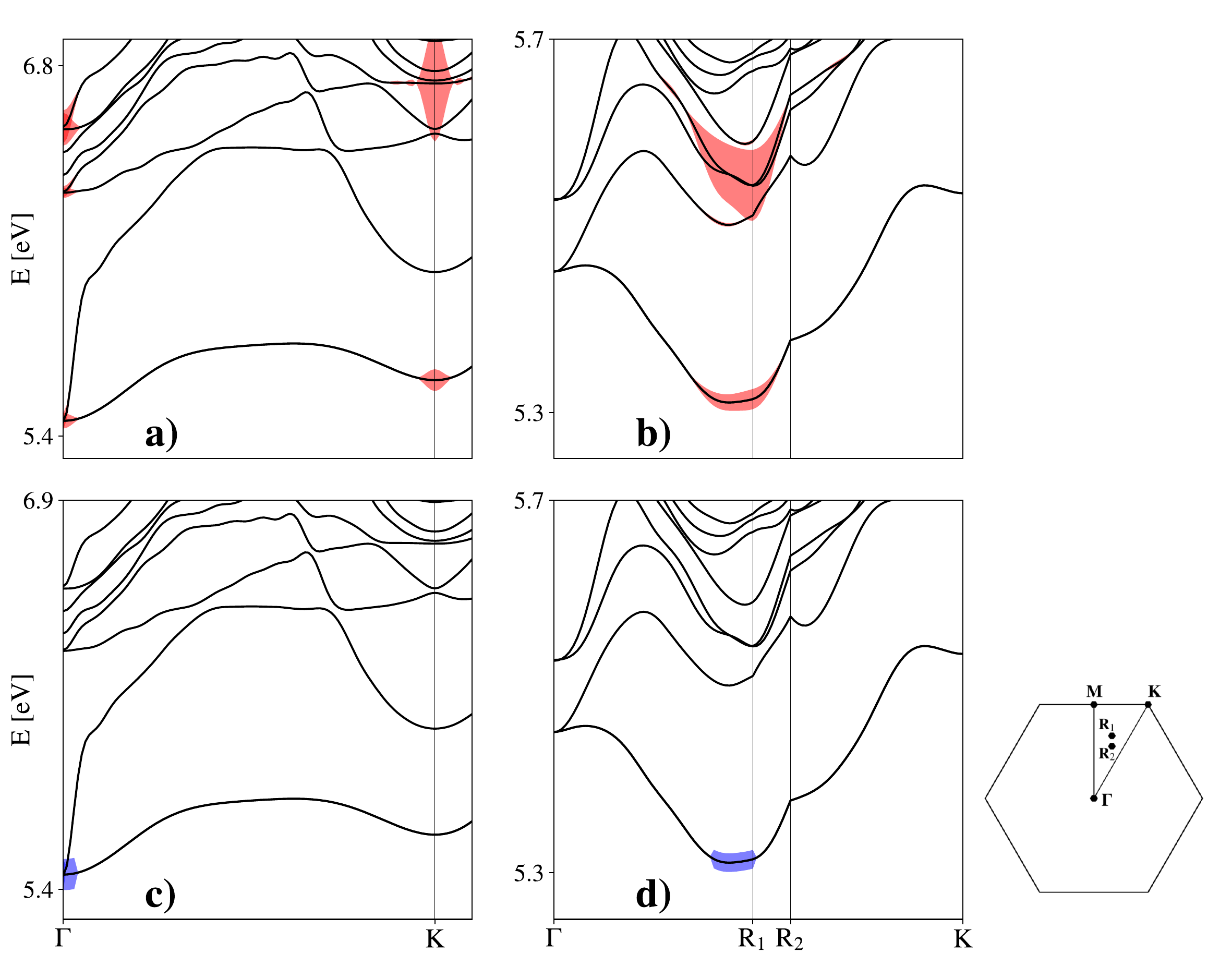}
	\caption{Details of the excitonic dispersions with occupation factors for both materials. Top row are rotated quasi-Fermi distributions, Eq.~\eqref{occ2} (red), bottom row are Boltzmann distributions, Eq.~\eqref{occ1} (blue). Panels a) and c) are monolayer hBN, while b) and d) show bulk hBN. The two types of occupation factors are evaluated at the same effective temperatures of $24$ K and $18$ K for monolayer and bulk, respectively. In the latter case, the high-symmetry line is modified to pass through the true minima of the bulk exciton dispersion according to our calculations. These are the points $R_1=(\tfrac{1}{9},\tfrac{2}{9},0)$ and $R_2=(\tfrac{1}{9},\tfrac{5}{18},0)$. Their location is shown in the inset to the right.\label{all_occup}}
\end{figure}

In the luminescence formula in the main manuscript excitons are occupied using a Boltzmann factor\cite{paleari2019exciton,paleari2019first}:
\be\label{occ1}
N_\lambda=e^{-\frac{E_{\lambda}-E_{min}}{kT_{exc}}}.
\ee
In this section we decided also to test the excitonic occupations used in Refs.~\onlinecite{Pedro2016PL,cannuccia2019theory}. In this formulation the excitonic occupations are coming from the nonequilibrium Green's function formalism, and consist in a rotation of the electronic quasi-Fermi distributions\cite{schafer2013semiconductor} in the exciton basis:
\bea\label{occ2}
N_\lambda&=&\sum_{t}  \langle \lambda | t  \rangle f^<_{K_1} \langle t | \lambda \rangle \label{fexc}, \\
f^{<}_K &=& f_{c \kk} (1 - f_{v \kk} )
\eea
where $t = \{v,c,\kk\}$. In Fig. \ref{all_occup} we compare the two types of occupation factor, Eqs.\eqref{occ1} and \eqref{occ2} for the two materials studied in our work.
We notice that Eq.~\eqref{occ2} produces wrong occupations of higher-energy excitons. 
This can be seen from Fig.~\ref{all_occup} (top panels), in which, in addition to the low-energy minima of the excitonic dispersion, also some exciton bands sitting $0.5\sim 1$ eV appear occupied in both hBN and m-hBN.
The corresponding occupations obtained with a Boltzmann distribution are shown in the bottom panels. 
While this does not matter when considering only a few excitonic bands\cite{cannuccia2019theory} or the $\Gamma$ point only\cite{libbi2022phonon}, it fails in more general situations like ours, with many excitons and a full $\qq$-point dispersion. It gives rise to unphysical luminescence spectra not reported here. For this reason we decided not to continue with Eq.~\eqref{occ2} any further.
The reason of this behavior lies in the fact that many exciton states lying at different energies are actually composed mostly of the same near-bandgap single-particle transitions. In the case of BN, it means that we consider the occupations of the $\pi$ and $\pi^*$ bands along the MK line in the BZ (plus around the H point for bulk): these correspond to the values of $t$ where the $\langle\lambda|t\rangle$ coefficient is non-negligible. If the excitonic occupations are written solely in terms of those transitions -- without a cutoff factor relative to the exciton energy landscape rather than just the electronic chemical potentials -- then nothing stops high-energy excitons to acquire finite occupation values.

\FloatBarrier
\bibliography{phassisted}
\nolinenumbers